\documentclass[apj,numberedappendix]{emulateapj}
\usepackage{amsfonts}
\usepackage{amsmath}
\usepackage{graphicx}

\def\so{1}
\def\sotxt{Steward Observatory, University of Arizona, 933 N. Cherry Avenue, Tucson, AZ 85721, USA; rskibba@ucsd.edu}
\def\stsci{13}
\def\stscitxt{Space Telescope Science Institute, 3700 San Martin Drive, Baltimore, MD 21218, USA}
\def\cea{12}
\def\ceatxt{AIM, CEA/Saclay, L'Orme des Merisiers, 91191 Gif-sur-Yvette, France}
\def\toulouseuni{6}
\def\toulouseunitxt{Universit\'{e} de Toulouse, UPS-OMP, IRAP, Toulouse, France}
\def\toulousecnrs{7}
\def\toulousecnrstxt{CNRS, IRAP, 9 Av.\ colonel Roche, BP 44346, F-31028 Toulouse cedex 4, France}
\def\strasbourguni{8}
\def\strasbourgunitxt{Universit\'{e} de Strasbourg, Observatoire Astronomique de Strasbourg, F-67000 Strasbourg, France}
\def\strasbourgcnrs{9}
\def\strasbourgcnrstxt{CNRS, Observatoire Astronomique de Strasbourg, UMR7550, F-67000 Strasbourg, France}
\def\cambridge{11}
\def\cambridgetxt{Institute of Astronomy, University of Cambridge, Madingley Road, Cambridge CB3 0HA, UK}
\def\missouri{14}
\def\missouritxt{314 Physics Building, Department of Physics and Astronomy, University of Missouri, Columbia, MO 65211, USA}
\def\leiden{10}
\def\leidentxt{Sterrewacht Leiden, Leiden University, PO Box 9513, 2300 RA Leiden, The Netherlands}
\def\wisconsin{5}
\def\wisconsintxt{Department of Astronomy, 475 North Charter St., University of
Wisconsin, Madison, WI 53706, USA}
\def\chile{15}
\def\chiletxt{Departamento de Astronomia, Universidad de Chile, Casilla 36-D, Santiago, Chile}
\def\iap{16}
\def\iaptxt{UPMC-CNRS UMR7095, Institut d'Astrophysique de Paris, F-75014 Paris, France}
\def\keele{17}
\def\keeletxt{Lennard-Jones Laboratories, Keele University, ST5 5BG, UK}
\def\princeton{4}
\def\princetontxt{Princeton University Observatory, Peyton Hall, Princeton, NJ 08544, USA}
\def\ray{3}
\def\raytxt{Raytheon Company, 1151 East Hermans Road, Tucson, AZ 85756, USA}
\def\ucsd{2}
\def\ucsdtxt{Center for Astrophysics and Space Sciences, Department of Physics, University of California, 9500 Gilman Dr., La Jolla, San Diego, CA 92093, USA}

\slugcomment{}

\shorttitle{Dust and Stellar Emission of the Magellanic Clouds}
\shortauthors{R. A. Skibba, C. W. Engelbracht, et al.}
\journalinfo{Accepted to the Astrophysical Journal on xx xx xxxx}

\begin{document}

%\title{The Emission by Dust and Stars of the Magellanic Clouds}% in the HERITAGE Survey
\title{The Spatial Distribution of Dust and Stellar Emission of the Magellanic Clouds}

\author
 {
  Ramin~A.~Skibba\altaffilmark{\so,\ucsd}, Charles~W.~Engelbracht\altaffilmark{\so,\ray}, 
  Gonzalo~Aniano\altaffilmark{\princeton}, Brian~Babler\altaffilmark{\wisconsin}, Jean-Philippe~Bernard\altaffilmark{\toulouseuni,\toulousecnrs}, Caroline~Bot\altaffilmark{\strasbourguni,\strasbourgcnrs}, Lynn~Redding~Carlson\altaffilmark{\leiden}, Maud~Galametz\altaffilmark{\cambridge}, Fr\'{e}d\'{e}ric~Galliano\altaffilmark{\cea}, Karl~Gordon\altaffilmark{\stsci}, Sacha~Hony\altaffilmark{\cea}, Frank~Israel\altaffilmark{\leiden}, Vianney~Lebouteiller\altaffilmark{\cea}, Aigen~Li\altaffilmark{\missouri}, Suzanne~Madden\altaffilmark{\cea}, Margaret~Meixner\altaffilmark{\stsci}, Karl~Misselt\altaffilmark{\so}, Edward~Montiel\altaffilmark{\so}, Koryo~Okumura\altaffilmark{\cea}, Pasquale~Panuzzo\altaffilmark{\cea}, Deborah~Paradis\altaffilmark{\toulouseuni,\toulousecnrs}, Julia~Roman-Duval\altaffilmark{\stsci}, M\'{o}nica~Rubio\altaffilmark{\chile}, Marc~Sauvage\altaffilmark{\cea}, Jonathan~Seale\altaffilmark{\stsci}, Sundar~Srinivasan\altaffilmark{\iap}, Jacco~Th.~van~Loon\altaffilmark{\keele} 
% and Remy Indebetouw?
 }

\altaffiltext{\so}{\sotxt}
\altaffiltext{\ucsd}{\ucsdtxt}
\altaffiltext{\ray}{\raytxt}
\altaffiltext{\princeton}{\princetontxt}
\altaffiltext{\wisconsin}{\wisconsintxt}
\altaffiltext{\toulouseuni}{\toulouseunitxt}
\altaffiltext{\toulousecnrs}{\toulousecnrstxt}
\altaffiltext{\strasbourguni}{\strasbourgunitxt}
\altaffiltext{\strasbourgcnrs}{\strasbourgcnrstxt}
\altaffiltext{\leiden}{\leidentxt}
\altaffiltext{\cambridge}{\cambridgetxt}
\altaffiltext{\cea}{\ceatxt}
\altaffiltext{\stsci}{\stscitxt}
\altaffiltext{\missouri}{\missouritxt}
%\altaffiltext{\ssctoulouse}{\ssctoulousetxt}
\altaffiltext{\chile}{\chiletxt}
\altaffiltext{\iap}{\iaptxt}
\altaffiltext{\keele}{\keeletxt}

%\newcommand{\bm}[1]{{\mbox{\boldmath $#1$}}} 
%\newcommand\nodata{ ~$\cdots$~ }

%\newcounter{appfig}
%\addtocounter{counter}{value}

%\maketitle


%\label{firstpage}

\begin{abstract}
\setcounter{footnote}{17}
We study the emission by dust and stars in the Large and Small 
Magellanic Clouds, a pair of low-metallicity nearby galaxies, 
as traced by their spatially resolved spectral energy 
distributions (SEDs).  This project combines \textit{Herschel} Space Observatory PACS and SPIRE far-infrared 
%photometry\footnotemark[14]
photometry\footnote{ The data were obtained as part of the HERschel Inventory of The Agents of Galaxy Evolution survey (HERITAGE; P.I., M. Meixner).} 
with other data at infrared and optical wavelengths.  
We build maps of dust and stellar luminosity and mass of both Magellanic Clouds, 
and analyze the spatial distribution of dust/stellar luminosity and mass ratios.  
These ratios vary considerably throughout the galaxies, generally between the range 
$0.01\leq L_{\rm dust}/L_\ast\leq 0.6$ and $10^{-4}\leq M_{\rm dust}/M_\ast\leq 4\times10^{-3}$.  
We observe that the dust/stellar ratios depend on the interstellar medium (ISM) environment, 
such as the distance from currently or previously star-forming regions, and on 
the intensity of the interstellar radiation field (ISRF). %radiative enviro, local heating sources 
%We also find that dust temperature, an indicator of 
%the heating of dust grains, is correlated more with 
%$L_\mathrm{dust}/M_\mathrm{dust}$ than with $L_\mathrm{dust}$ alone, implying that 
%the dust emission per unit mass is closely related to how much dust is heated. 
In addition, we construct star formation rate (SFR) maps, and find that the SFR 
is correlated with the dust/stellar luminosity and dust temperature in both 
galaxies, demonstrating the relation between star formation, dust emission and 
heating, though these correlations exhibit substantial scatter.
% add something about SFE \& dust/gas ratios?
\end{abstract}


%% 12) Any footnotes in the paper title should be set as
%% \title{Title\footnotemark[1]} \footnotetext[1]{text} (with foootnotetext
%% outside the title), rather that simply a \footnote{}. 
%%
%% NOTE that if you use the footnote in the title, the footnote counter for
%% the main text will be wrong. You need to reset it manually _after_ the
%% first \section. For example, if the frontmatter footnotes (title +
%% affiliations) stop at 3, you need \setcounter{footnote}{3}

\section{Introduction}\label{sec:intro}

\setcounter{footnote}{18}

Among the nearby galaxies, the Small Magellanic Cloud (SMC) and Large Magellanic 
Cloud (LMC) represent unique astrophysical laboratories for studies of the 
lifecycle of the interstellar medium (ISM), because of their proximity 
($\approx 50$ and $60~\mathrm{kpc}$ for the LMC and SMC, respectively; Ngeow \& Kanbur 2008; Szewczyk et al.\ 2009), %Feast '99; Keller & Wood '06
low metallicity ($\approx1/5$ and $1/2~Z_\odot$; Dufour et al.\ 1982; Russell \& Dopita 1992), 
and favorable viewing angle ($24^\circ$ for the LMC; Nikolaev et al.\ 2004). %2008?%van der Marel \& Cioni 2001  
The SMC and LMC (MCs) are well-suited for detailed studies of galaxy evolution processes, 
either as whole galaxies or on the scales of individual star-forming regions or resolved stellar populations. 
In particular, these galaxies are suited for studies of the ISM phases, interstellar dust properties, and star formation rates as a function of metallicity and radiative environment (i.e., the intensity and hardness of the radiation field). 

%mention perspective of Magellanic Clouds as `satellite' galaxies 
The MCs are also unusual as relatively bright `satellite' galaxies (compared to other satellites) 
very close to the Milky Way (MW). %, the `central' galaxy of the Local Group. 
The luminosity distributions of satellite galaxies in groups have been the subject 
of many statistical studies (e.g., Skibba et al.\ 2007; van den Bosch et al.\ 2007), %Paranjape \& Sheth 2012
and some recent studies have focused on MW-like galaxies with MC-like satellites in 
the Sloan Digital Sky Survey, Galaxy And Mass Assembly, and in models and simulations (e.g, Liu et al.\ 2011; Tollerud et al.\ 
2011; Sales et al.\ 2011; Busha et al.\ 2011; Robotham et al.\ 2012). %and maybe Starkenburg et al. (2012) 
% also maybe Lokas et al. (2011, arXiv:1112.5336)
%maybe Tsujimoto T., Bekki K., 2009, ApJ, 700, L69; or Bekki (2008); or Bekki \& Chiba (2009)
%maybe Gallart et al., 2008, ApJ, 682, L89
% could also mention van den Bergh (1999)
In addition, analyses of the MCs' star formation, structures, and velocities 
%including distance and proper motion measurements, 
have yielded evidence for tidal interactions between the LMC and SMC: 
%for example, 
the Magellanic Bridge and SMC Wing %Karl's term is "Tail" 
appear to consist of tidally stripped material; there is tidally 
triggered star formation within the MCs and the Magellanic Bridge; and the LMC's 
offset and warped bar are arguably due to a previous collision with the SMC %and van der Marel 2001
(Yoshizawa \& Noguchi 2003; Mastropietro et al.\ 2005; Harris 2007; D'Onghia \& Lake 2008; 
Gordon et al.\ 2009, 2011; Bekki 2011; Besla et al.\ 2012; Cignoni et al.\ 2012). %also No\"{e}l et al. 2009; McCumber et al. 2005; Besla et al. 2012; Barger et al., in prep. 
% D'Onghia E., Lake G., 2008, ApJ, 686, L61; see also Mastropietro et al. (2005)
% Bekki K., 2011, MNRAS, 416, 2359 and Bekki K., Chiba M., 2007, PASA, 24, 21
%
%some have speculated about a possible interaction with the MW as well, though the MCs' structural properties make this unlikely (e.g., de Vaucouleurs \& Freeman 1972; Besla et al.\ 2012). 
% also could mention Besla et al. (2007); Choi, Weinberg & Katz (2009) or Weinberg (2000)

Nearly half of the bolometric luminosity of the Universe is channeled through the 
mid- and far-infrared (IR) emission of galaxies (e.g., Hauser \& Dwek 2001). 
%This spectral region also probes important physical properties of galaxies, such 
%as their metal content, dust content, and cold gas content (e.g., Draine et al.\ 2007). 
Therefore, in order to develop a more complete picture of galaxy evolution, 
it is necessary to study the IR emission that is reprocessed by dust in galaxies. 
We seek to understand the physical processes that regulate galaxy evolution, 
including the formation of stars and the interstellar radiation field (ISRF), 
and the return of radiant energy from these stars into the ISM. 

%dust analyses of MCs, constraints on dust mass and emission (Gordon, Bernard, Galliano, Paradis)
IR observations of galaxies have yielded insight into the nature and spatial distribution 
of their dust content, including its extinction and emission properties (e.g., dust temperature), 
as well as its composition, grain size distribution, and abundance (e.g., dust mass). 
%IR observations of galaxies have yielded insight into their 
%dust properties, including the emission, extinction, spatial distribution, 
%size distribution, abundance, (including PAHs, VSGs, BGs), 
%mass, and temperature of dust grains.  
A variety of studies have investigated these 
issues in the MCs, often utilizing mid- and far-IR observations, in order to 
constrain their dust properties (e.g., % and Li \& Draine 2002; 
Weingartner \& Draine 2001; Bot et al.\ 2004, 2010; Sakon et al.\ 2006; 
Bernard et al.\ 2008; Paradis et al.\ 2009, 2011; Gordon et al.\ 2010; 
Sandstrom et al.\ 2010; Galliano et al.\ 2011; Planck Collaboration 2011). 
%maybe mention issue of submillimeter excess? %but we discuss it later! %(Dumke et al.\ 2004; Bot et al.\ 2010; Galametz et al.\ 2009, 2011; Gordon et al.\ 2010; Galliano et al.\ 2011; Planck Collaboration et al. 2011). %also Gordon et al., in prep.; Hunt et al., in prep.
% Dumke M., Krause M., Wielebinski R., 2004, A\&A, 414, 475
% and Galliano F., et al., 2005, A\&A, 434, 867
% and Zhu M., Papadopoulos P. P., Xilouris E. M., Kuno N., Lisenfeld U., 2009, ApJ, 706, 941 
% also Draine B.T., Hensley B., 2012, ApJ, submitted
By combining IR and other data, some studies have also explored how dust 
production, destruction, and heating are related to atomic and molecular 
gas content, star formation, and the interstellar radiation field. 
For example, constraints on the dust and gas content are sufficient to 
estimate gas-to-dust ratios and deviations from its expected metallicity 
dependence (Stanimirovi\'{c} et al.\ 2000; Gordon et al.\ 2003; Leroy et al.\ 
2007; Galliano et al.\ 2011). % Roman-Duval et al., in prep.). 
%\textbf{In addition, star catalogs and IR emission have allowed for analyses of the dust and stellar populations in the Magellanic Clouds (Sauvage et al.\ 1990; Zaritsky 1999).}

In general, analyses of the spectral energy distributions (SEDs) of galaxies 
yield information about the relation and balance between dust and stellar 
emission, by exploiting multi-wavelength observations (e.g., Draine et al.\ 
2007; Dale et al.\ 2007; da Cunha et al.\ 2010; Skibba et al.\ 2011). 
\textit{The spatial distribution of dust vis-\`a-vis stellar luminosity and mass is the main focus of this paper.}  
The distribution of dust/stellar ratios within galaxies, and its connection to 
star formation and dust heating, has important implications about the energy conversion from stellar to dust emission. %balance of dust and stellar emission.
%Julia Our goal is to explore the energy balance abundance of dust and stars throughout the LMC and SMC and their implications for galaxy evolution.  talk about why it's important to know about the energy balance, stellar/dust ratio in galaxy evolution.

It is now possible, with data from \textit{IRAS}, \textit{Spitzer}, and \textit{Herschel}, to constrain the radial distributions of dust and stars within 
galaxies (Sauvage et al.\ 1990; Zaritsky 1999; Mu\~{n}oz-Mateos et al.\ 2009; Mattsson \& Andersen 2012), %Hunt et al., in prep.), 
and to analyze the spatial distribution of dust, gas, and starlight in the  
MW (Paradis et al.\ 2012), M31 (Smith et al.\ 2012), M33 (Boquien et al.\ 2011), and M83 (Foyle et al.\ 2012), 
and other nearby galaxies (Aniano et al.\ 2012; Galametz et al.\ 2012). 
%in addition, there are related studies of stars and HII regions of the LMC/SMC, such as Harris \& Zaritsky; also Bell et al.\ (2002), Blair et al.\ (2009)... %also see Slater et al., Paradis et al. 
Others have added optical, $H\alpha$, and UV observations to constrain the 
evolution and environmental dependence of star formation in the LMC and SMC 
(Bell et al.\ 2002; Harris \& Zaritsky 2004, 2009; 
Blair et al.\ 2009; Lawton et al.\ 2010), and clustering analyses have 
demonstrated the hierarchical formation of star clusters and dust clumps (Bonatto \& Bica 2010; Kim et al.\ 2010). 

% maybe add emphasis?
We build upon this work by examining the resolved spatial distribution of dust 
and stellar emission within the MCs, using new far-IR and 
submillimeter (submm) photometry from the HERITAGE survey (Meixner et al.\ 2010; Meixner et al., in prep.), 
near-IR photometry from the \textit{Spitzer} Surveying the Agents of 
Galaxy Evolution (SAGE, Meixner et al.\ 2006; SAGE-SMC, Gordon et al.\ 2011) surveys, and 
%$H\alpha$ emission surveys (SHASSA, MCELS). %Gaustad et al. 2001, Smith et al. 1999 
the Southern $H\alpha$ Sky Survey Atlas (SHASSA, Gaustad et al.\ 2001). 
Our goal is to explore the energy balance and abundance of dust and stars 
throughout the LMC and SMC, and their implications for galaxy evolution.

This paper is organized as follows. 
We describe the data and their processing in the next section, and in Section~\ref{sec:description}, we describe the dust and stellar properties that we analyze in the paper. 
In Sections~\ref{sec:results}, \ref{sec:dusttostellar}, and \ref{sec:sfr}, we present our main results: we construct and interpret maps of dust luminosity, dust/stellar luminosity and mass, and SFR, for the LMC and SMC. 
We focus on the relation between dust/stellar ratios and other properties, especially dust heating and star formation. 
%We then discuss differences between the Magellanic Clouds and selected regions within them in Sections~\ref{sec:diffs} and \ref{sec:regions}, and 
We conclude with a summary and discussion of our results.
%we construct and interpret maps of dust luminosity, and describe temperature and mass maps. 
%This is followed by the construction of stellar luminosity and mass maps, and an 
%analysis of the distribution of dust/stellar 
%luminosity and mass in Section~\ref{sec:dusttostellar}. 
%We also construct maps of star formation rate (SFR) in Section~\ref{sec:sfr}, 
%and examine the relation between SFR and dust heating. 
%These two sections constitute the most important results of the paper. 
%We then discuss differences between the Magellanic Clouds and selected 
%regions within them in Sections~\ref{sec:diffs} and \ref{sec:regions}. 
%We conclude with a summary and discussion of our results. 

\section{Data}\label{sec:data}

We briefly describe below the data used in this paper.  We refer the reader to Meixner et al.\ (2010; Meixner et al., in prep.) for details about the observations using the \textit{Herschel} Space Observatory (Pilbratt et al.\ 2010) and data reductions. 

%[First describe the PACS (Poglitsch et al.\ 2010) and SPIRE (Griffin et al.\ 2010) photometry and data reduction. We refer the reader to Meixner et al.\ (in prep.) for details.] 
HERITAGE is a uniform survey of the LMC, SMC, and Magellanic Bridge with the Spectral and Photometric Imaging Receiver (SPIRE) at 250, 350 and 500~$\mu$m (Griffin et al.\ 2010) and the Photodetector Array Camera and Spectrometer (PACS) at 100 and 160~$\mu$m (Poglitsch et al.\ 2010). 
The SPIRE beam sizes are approximately 18, 25, and $37''$ at 250, 350, and 500~$\mu$m, and the flux calibration uncertainties are at most $15\%$. 
The PACS beam sizes are 7 and $11''$ at 100 and 160~$\mu$m, and the calibration uncertainties are at most $20\%$. 
The data were processed using the HIPE 7.0 data reduction software (Ott 2010).
The images were converted from Jy/pixel (PACS) and Jy/beam (SPIRE) to MJy/sr.
% The PACS and SPIRE calibration uncertainties are 10\% and 7\%, respectively. %right?

%Then briefly describe the MIPS and IRAC data (Meixner et al.\ 2006; Gordon et al.\ 2011).  
%Infrared Array Camera (IRAC) and Multiband Imaging Photometer (MIPS); cite Werner+ for Spitzer
To these we add ancillary data, including imaging from the MIPS (Rieke et al.\ 2004; 24, 70, and 160~$\mu$m) 
and IRAC (Fazio et al.\ 2004; 3.6, 4.5, 5.8, and 8~$\mu$m) instruments on board \textit{Spitzer} Space Telescope (Meixner et al.\ 2006; Gordon et al.\ 2011), 
and $H\alpha$ imaging from the SHASSA (Gaustad et al.\ 2001) %and MCELS (Smith et al.\ 1999) 
survey.
%then whatever optical, $H\alpha$, and/or UV data we choose to use. %(which are at poorer resolution).

%500um: 36.3" beam FWHM, 14" pixel size (3.4 or 4.1 pc for the MCs)
The LMC and SMC multi-wavelength photometry are at different resolutions.  In order to use these data together, it is necessary to convolve them to a common %the poorest, SPIRE $500\mu$m, 
resolution (that of SPIRE $500~\mu$m) to generate images with a common point spread function (PSF).  
We use the convolution kernels of Aniano et al.\ (2011), which use techniques similar to those of Gordon et al.\ (2008). 
%
% Julia: need more details about the background subtraction
We then subtract a constant background from the images, by fitting Gaussian distributions to the flux densities beyond the galaxies; data within $1\sigma$ of the background are not used. 
We subsequently crop the images and align them with each other,  
yielding image arrays of the same dimensions, with $14\prime\prime$ pixel size (3.4 and 4.1~pc linear size for the LMC and SMC). 
These are the images that we use for computing dust and stellar luminosities and masses 
in the following sections.

%%[example IR SEDs of a couple regions (LMC's bar; N11 (Walborn \& Parker 1992) in the NW spiral arm; NW void (similar to HZ09 def., east of N11 and 5x larger), which is ~quiescent; region centered at NGC~346, a star-forming region in the SMC (Simon et al.\ 2007)).]
%For an example, we show the IR SEDs of a few regions within the Magellanic Clouds in Figure~\ref{fig:SEDs}: the LMC's bar; the star-forming region N11 (Walborn \& Parker 1992), in the LMC's NW spiral arm; the LMC's `NW void' (defined similarly to Harris \& Zaritsky 2009), east of N11 and five times larger; and NGC~346, a star-forming region in the SMC (Simon et al.\ 2007).  
%Note that the flux scales are different in each panel of the figure.
%\begin{figure}
% \includegraphics[width=\hsize]{LMCandSMC_4regionSEDs.png} 
% \caption{Infrared SEDs of regions within the LMC (the bar, upper left; N11, upper right; and the `NW void', lower right) and SMC (NGC~346, lower left).  Note that the flux scales are different in each panel.}
% \label{fig:SEDs}
%\end{figure}

\section{Description of Galaxy Properties}\label{sec:description}

In this section, we describe the dust and stellar properties of the Magellanic Clouds that we will use throughout the paper.  Our goal is to analyze the relations between these dust and stellar properties within the MCs.  All of the properties are inferred from the resolved SEDs, except for the SFR, which includes $H\alpha$ data.

\subsection{Dust Properties}

\subsubsection{Dust Luminosity}\label{sec:Ldustdesc}

The `dust luminosity' is a useful quantity 
because it can be directly inferred from the IR fluxes, and because it can be 
used as a proxy for the obscured star formation as well as the temperature of 
dust grains (e.g., Dale \& Helou 2002; Draine \& Li 2007).

%We begin by defining a quantity that we refer to as the dust flux or luminosity. 
%This ``total'' luminosity, which is calibrated at wavelengths at or shorter than $\lambda=160\mu$m, 
%can be compared to what we refer to as a dust flux or luminosity. %Eqn.~\ref{eq:fdust} 
Following Skibba et al.\ (2011; hereafter S11), we measure the dust luminosity $L_\mathrm{dust}$ 
as follows: 
\begin{equation}
%  \frac{f_\mathrm{dust}}{f_\ast} \,\equiv\,
%    \frac{ \int_{\lambda=4.5\mu m}^{\lambda=850\mu m}\,{\rm d}\nu\, f_\nu \,-\, \int_{4.5\mu m}^{50\mu m}\,{\rm d}\nu\,f_\mathrm{star} } 
%         { \int_{\lambda=0.15\mu m}^{\lambda=4.5\mu m}\,{\rm d}\nu\, f_\nu \,+\, \int_{4.5\mu m}^{50\mu m}\,{\rm d}\nu\,f_\mathrm{star} } \, , 
 L_\mathrm{dust} \,\equiv\, 
   %4\pi D^2\,\int_{\lambda=\mathrm{IRAC}5.8\mu m}^{\lambda=\mathrm{SPIRE}500\mu m}\, f_\nu\,{\rm d}\nu, 
   %4\pi D^2\,\int_{\nu_\mathrm{min}}^{\nu_\mathrm{max}}\, f_\nu\,{\rm d}\nu, 
   4\pi D^2\,\int_{\lambda_\mathrm{min}}^{\lambda_\mathrm{max}}\, f_\lambda\,{\rm d}\lambda, 
 \label{eq:fdust}
\end{equation}
%where ${\rm d}\nu={\rm d}(\mathrm{log}\nu)c/\lambda$. 
where $D$ is the distance to the galaxy, $f_\lambda$ is the flux density at wavelength $\lambda$, 
and the integration is performed between 
$\mathrm{IRAC}~5.8\mu{\rm m}\leq\lambda\leq\mathrm{SPIRE}~500\mu{\rm m}$. 
The resolved SED is interpolated between the bands over this range, and pixels with noisy or missing far-IR data are not included. 
Unlike in S11, we have not subtracted the stellar emission from the mid-IR bands before performing the integration, as they found that the level of contamination was negligible for dwarf and late-type galaxies (less than a few per cent); the contamination by starlight is negligible throughout the LMC and SMC as well. 
We also do not extrapolate beyond 500$\mu$m, as this contribution is negligible as well (less than $0.5\%$).  $L_\mathrm{dust}$ is usually dominated by the fluxes between $70\mu{\rm m}\leq\lambda\leq250\mu{\rm m}$. 

%We express both luminosities in units of erg/s/sr. \textbf{[update this.]} 
The luminosities are converted from erg/s/sr to erg/s/${\rm pc}^2$, 
%by accounting for the SPIRE500$\mu$m pixel size, 
and then we normalize by the solar luminosity $L_\odot$. 
%\textbf{Shall we extrapolate $L_\mathrm{dust}$ to longer wavelengths, and try to subtract stellar component, like we did in the KINGFISH paper?} 
By summing over the pixels of these $L_\mathrm{dust}$ surface density maps, we obtain total luminosities of $2.5\times10^7~L_\odot$ for the SMC and $2.3\times10^8~L_\odot$ for the LMC.  These values appear to be consistent with the SED analyses of Israel et al.\ (2010) and Meixner et al.\ (in prep.), though a direct comparison is not possible.

\subsubsection{Dust Temperature}\label{sec:Tdustdesc}

%The dust temperatures and masses are estimated from modified blackbody fitting and are taken from Gordon et al.\ (in prep.), which uses an approach similar 
%to that of Engelbracht et al.\ (2010), Gordon et al.\ (2010), and S11. 
%The temperatures are computed by fitting the SED, 
%using the PACS and SPIRE far-IR and submm ($100-500\mu$m) flux densities, 
%to a single temperature blackbody modified by an emissivity power-law (Hildebrand 1983): 
The dust temperatures and masses are estimated from 
fitting the PACS and SPIRE far-IR and submm (100--500\,$\mu$m) SED
with a single-temperature blackbody
(i.e., a blackbody of temperature $T_{\rm dust}$) 
modified by an emissivity law of $\lambda^{-\beta}$: 
\begin{equation}
 f_\lambda \propto \lambda^{-\beta} B_\lambda(T_\mathrm{dust}) ,
\label{modBB}
\end{equation}
where $B_\lambda$ is the Planck function. %, which is $2ckT/\lambda^4$ in the Rayleigh-Jeans limit. 
%%I'm only using these modBB fits for the 70um excess, after all.
%This is an approach similar to that of Engelbracht et al.\ (2010), Gordon et al.\ (2010), and S11, who also applied this method to nearby galaxies. 
%In this case, a $\beta=1.5$ emissivity law is assumed at $\lambda<300\mu$m, while at longer wavelengths the emissivity law is allowed to vary to fit the surface brightnesses measured in the SPIRE 350 and 500$\mu$m bands. 
The emissivity exponent at $\lambda<300~\mu$m is taken to be $\beta=1.5$. 
%The emissivity law at $\lambda<300\mu$m assumes a $\beta=1.5$ emissivity law, 
At longer wavelengths, the emissivity law is allowed to vary so as to fit the surface brightnesses 
measured in the SPIRE 350 and 500~$\mu$m bands, and typically varies between $1<\beta<2$ throughout the galaxies.  
These values of the emissivity index are consistent with previous studies (Sakon et al.\ 2006; Paradis et al.\ 2011; Planck Collaboration 2011). 
This model has been developed to produce small fractional residuals and to be consistent with constraints from gas-to-dust ratios (see also Galliano et al.\ 2011). 
%For details, we refer the reader to Gordon et al.\ (2010) \textbf{and Gordon et al.\ (in prep.)}, where the dust temperature and mass maps of the MCs are presented. 
We use preliminary dust temperature (and mass) maps from Gordon et al.\ (in prep.), and we refer the reader to that paper and Gordon et al.\ (2010) for details. 
The approach here is similar to that of Engelbracht et al.\ (2010) and S11. 

We note that the dust temperature $T_{\rm dust}$ and the emissivity index $\beta$ are not independent; indeed, 
an inverse correlation between $T_{\rm dust}$ and $\beta$ has been observed, 
which some have argued is due to physical properties of the dust grains (Dupac et al.\ 2003; Paradis et al.\ 2011; Juvela \& Ysard 2012) or to noise (Shetty et al.\ 2009; Kelly et al.\ 2012). %and maybe also mention Deborah's TLS model.
%The varying emissivity law to some extent does account for the submm excess 
%inferred by some authors, which may be due to a very cold dust component, 
%though its origin is still under debate (e.g., Gordon et al.\ 2010). 

%\textbf{[mention Appendix~\ref{app:70umexcess}, in context of modBB fitting.]}
The modified-blackbody approach closely reproduces the 100--500\,$\mu$m 
PACS and SPIRE photometry, but underpredicts the MIPS 70\,$\mu$m photometry.
The `70~$\mu$m excess' (i.e. the excess of the 70\,$\mu$m flux densities measured 
by MIPS over that predicted from the single-temperature modified-blackbody fitting) 
is briefly discussed in Appendix~\ref{app:70umexcess} as arising from 
stochastically heated small grains, or from the fact that 
the single-temperature assumption is not realistic. %: the grains of different sizes have different temperatures when exposed to starlight of different intensities. 
%and/or the fact that the dust grains in the Magellanic Clouds
%are actually heated to a range of temperatures 
%(instead of a single temperature).
%In Appendix~\ref{app:70umexcess}, we briefly discuss the `70$\mu$m excess' (Bernard et al.\ 2008), 
%which is estimated using the results of modified blackbody fitting and is a measure of stochastically heated dust grains.

\subsubsection{Dust Mass}\label{sec:Mdustdesc}

%\textbf{[need to revise this, since now using $M_\mathrm{dust}$ of Gordon et al. (in prep.).]} 
%Following Gordon et al.\ (in prep.), 
With $T_\mathrm{dust}$ (and $\beta$) determined from Eqn.~(\ref{modBB}), 
we obtain the dust mass $M_{\rm dust}$ from $f_\lambda$:
\begin{equation}
  M_\mathrm{dust} \,=\, \frac{f_\lambda\,4\pi D^2}
                             {\kappa_{\mathrm{abs},\lambda}\,4\pi B_\lambda(T_\mathrm{dust})}
 \label{Mdust}
\end{equation}
%where $f_\lambda$ is the flux density, $D$ is the distance from the galaxy, and the additional $4\pi$ factor in the denominator is due to integrating over steradians. 
%$\kappa_{\mathrm{abs},\lambda}$ is the mass absorption efficiency at a given wavelength (Laor \& Draine 1993; Li \& Draine 2001):  
%$\kappa_{160\mu{\rm m}}=Q_{\rm em}/((4/3)a\rho)$, where $a=0.1\mu$m is the grain radius, the grains are assumed to be spherical silicate grains with density $\rho=3~{\rm g}/{\rm cm}^{-3}$, and $Q_{\rm em}(160\mu{\rm m})=5.5\times10^{-4}$, resulting in $\kappa=13.75~{\rm cm}^2/{\rm g}$. 
where $\kappa_{\mathrm{abs},\lambda}$ is the mass absorption coefficient 
at wavelength $\lambda$. 
This is modeled as a single component with an average temperature.  A model including a second, cold dust component also yields good fits, though recent analyses of far-IR SEDs with Herschel data have shown that assuming a cold dust component yields dust masses that are too large based on metallicity constraints (e.g., Bot et al.\ 2010; De Looze et al.\ 2012; Aniano et al.\ 2012; Galametz et al.\ 2012). 

%Following Gordon et al.\ (in prep.), 
To compute $M_\mathrm{dust}$ in Eqn.~(\ref{Mdust}), for spherical grains of radius $a$,
$\kappa_{\mathrm{abs},\lambda} = 3\,Q_{\rm abs}(a,\lambda)/\left(4\,a\,\rho\right)$, 
where $Q_{\rm abs}(a,\lambda)$ is the absorption efficiency and $\rho$ is the mass 
density of the dust.  For submicron-sized silicate grains 
($\rho\approx 3\,{\rm g}\,{\rm cm}^{-3}$) $Q_{\rm abs}(160\,\mu{\rm m})\approx 5.5\times10^{-4}$,  
we obtain $\kappa_{\mathrm{abs},160\,\mu{\rm m}}
\approx 13.75\,{\rm cm}^2\,{\rm g}^{-1}$ (Laor \& Draine 1993; Li \& Draine 2001).
The masses are computed at $\lambda=160~\mu$m, %not 250um?
in order to minimize the dependence 
on temperature and emissivity variations, as well as flux uncertainties (Gordon et al., in prep.). 
In contrast, S11 used 500~$\mu$m as the reference wavelength; however, the 
relative spatial distribution is not strongly dependent on this choice. 
The absolute values of the masses may be affected by up to $0.2~{\rm dex}$ by the adopted reference wavelength (see S11), though for low-mass irregular galaxies with significant `submm excesses' (e.g., Bot et al.\ 2010; Galametz et al.\ 2011) it is accurate to use $160~\mu$m. 
%Following Skibba et al.\ (2011), we compute the masses at $\lambda=500\mu$m, 
%in order to minimize the dependence on the temperature, although 
%the uncertainties of the flux densities are larger than at shorter wavelengths. 
%At $500\mu$m, $\kappa_\mathrm{abs}=0.95\,\mathrm{cm}^2/\mathrm{g}$ 
%(which is $\approx20\%$ lower than the value quoted in Li \& Draine (2001)). 

By summing over the pixels of these dust mass (surface density) maps, we obtain 
total masses of $1.1\times10^6~M_\odot$ for the LMC and $1.1\times10^5~M_\odot$ for the SMC. 
% total dust masses: 1066091.8 (1.07e6) for LMC, 108786.26 (1.09e5) for SMC
The LMC mass estimate is consistent with the value of $1.2\times10^6~M_\odot$ obtained by Bernard et al.\ (2008) and Galliano et al.\ (2011), and the SMC mass is approximately consistent with $3\times10^5~M_\odot$ obtained by Leroy et al.\ (2007). 
The spatial resolution here is sufficient to yield unbiased total masses (see also Galliano et al.\ 2011; Aniano et al.\ 2012); based on the results of Galliano et al., the mass values may be affected by up to 10\%.

\subsection{Stellar Properties}

\subsubsection{Stellar Luminosity}\label{sec:Lstardesc}

We will use the 3.6~$\mu$m and 4.5~$\mu$m photometry as an indicator of stellar luminosity and mass in this section and the next section.  
Emission at these wavelengths is strongly correlated with the spatial distribution of stellar light and mass, especially that of old stars, but using these wavelengths alone could be contaminated by nonstellar contributions (Mentuch et al.\ 2010; Zibetti \& Groves 2011; Meidt et al.\ 2012). %cf. Sani et al.\ 2011) 
%maybe also mention Marble et al. (2010); see his Sec. 4.2.1
For example, hot dust and the 3.3~$\mu$m PAH (polycyclic aromatic hydrocarbon) emission can cause appreciable contamination, 
contributing $\sim20\%$ of the integrated light at 3.6~$\mu$m in star-forming regions (Meidt et al.\ 2012). %maybe also look at Povich et al. (2007)? see Lynn's comments.
The near-IR $H$-band luminosity would be a superior tracer of stellar light and mass (Zibetti et al.\ 2009), for example, 
%but currently such data are lacking for the Magellanic Clouds.%\footnote{shall we explain problems with available 2MASS data?  and AKARI is not a big improvement on IRAC here.}, we instead utilize 3.6 and 4.5$\mu$m data, which have the advantage of high angular resolution and small photometric errors. 
but such measurements with resolution and relative sensitivity similar to the
other data used here do not yet exist.
%2MASS: Jarrett T. H., Chester T., Cutri R., Schneider S. E., Huchra J. P., 2003, AJ, 125, 525; AKARI: Isihara D., et al., 2010, A\&A, 514, 1

% (25 May) Sacha suggested using extinction information, such as UV data or point sources or stellar flux density map from Harris & Zaritsky.  This would give us something more physical than just L3.6.  Or what I could do would be to use this information to constrain how I infer the total L* from L3.6 (or 3.6 & 4.5um). 

%[Attempt to estimate \textbf{bolometric stellar luminosity}: $L_{3.6\mu{\rm m}}\rightarrow L_\ast$, plus $L_{3.6}/L_{4.5\mu{\rm m}}$ color correction.]
Here we will attempt to infer a `bolometric' stellar luminosity $L_\ast$, comparable to the dust luminosity $L_\mathrm{dust}$ (Eqn.~\ref{eq:fdust}). 
For most of the nearby galaxies in S11, $L_\ast$ was essentially estimated from UV to 5~$\mu$m wavelengths, allowing us to constrain the distribution of $L_\ast$ as a function of $L_{3.6\mu{\rm m}}$ and $L_{3.6}/L_{4.5\mu{\rm m}}$ for galaxies similar to the LMC and SMC, with similar morphologies and masses.   
We can also estimate similar global quantities for the MCs, using the published SEDs (Israel et al.\ 2010; Gordon et al.\ 2011). 
The LMC and SMC have $L_\ast/L_{3.6\mu{\rm m}}\approx80$ and 130, respectively, which 
are slightly lower values than the comparable galaxies in S11 (which had $\sim100-300$), 
but their $L_{3.6}/L_{4.5\mu{\rm m}}$ ratios are similar.  
At fixed $3.6~\mu$m luminosity, it is unclear whether or to what extent $L_\ast$ is significantly correlated with $L_{3.6}/L_{4.5\mu{\rm m}}$, so we choose not to include a correction based on this. 

For simplicity, we adopt $L_\ast=fL_{3.6\mu{\rm m}}$, where $f$ is a proportionality constant based on the global stellar SEDs of the LMC and SMC (i.e., the factor of 80 or 130), while acknowledging that this approximation introduces an uncertainty of $\sim25\%$, based on the distribution of $L_\ast/L_{3.6\mu{\rm m}}$ of the comparable nearby galaxies in S11. %[check this]. 
In addition, from the 3.6~$\mu$m residual (point-source subtracted) images (Meixner et al.\ 2006; Gordon et al.\ 2011, we estimate that our $L_\ast$ values 
%residuals: ~0.1 MJy/sr on average 
are affected up to $10\%$ from contamination by non-stellar sources. 
The resulting $L_\ast$ maps have uncertainties of $\sim0.15$ dex.
% remember to update plots: *MC_LdustLstarbolconst_ratio_Karlnanbg.png maps
% and contour plots: LMCandSMC_SFRvsLdustLstar_contour.png & LMCandSMC_SFRvsTdust_contour_v2.png

\subsubsection{Stellar Mass}\label{sec:Mstardesc}

To estimate the MCs' stellar mass, we will use the calibration of Eskew et al.\ (2012). 
This calibration converts from the 3.6 and 4.5~$\mu$m flux densities ($f_{\rm 3.6\mu m}$ and $f_{\rm 4.5\mu m}$) to stellar mass $M_{\star}$, 
and it is based on a detailed analysis of the LMC, for which they find a scatter of approximately 30\% in mass around the mean relationship: 
\begin{equation}
  M_\ast \,=\, 10^{5.65}\,f_{3.6\mu{\rm m}}^{2.85}\,f_{4.5\mu{\rm m}}^{-1.85}\,
    \biggl(\frac{D}{0.05}\biggr)^2\,M_\odot , 
 \label{Mstar}
\end{equation}
where the flux densities $f_{\rm 3.6\mu m}$ and $f_{\rm 4.5\mu m}$ are in Jy and the distance $D$ is in Mpc.

As discussed above, using these wavelengths rather than near-IR (e.g., $H$-band) 
fluxes yields a potentially 
biased indicator of stellar mass, and may in some regions be affected by hot dust 
and PAH emission (e.g., Zibetti \& Groves 2011). 
In particular, Eskew et al.\ (2012) note that regions that deviate significantly from 
their calibration often have young stellar populations ($<300$ Myr old) and high 8~$\mu$m flux. 
They also acknowledge that the calibration could depend on metallicity, but they were 
unable to explore this issue because there is little variation in metallicity in 
the LMC (Pagel et al.\ 1978); however, constraining any metallicity dependence of the calibration is beyond the scope of this paper. 
Note that Eskew et al.\ adopt a Salpeter initial mass function (IMF), and argue that their inferred stellar mass-to-light ratios moderately disfavor a bottom-light IMF. 
The reader should bear these caveats in mind.

%\textbf{first comment on $M_\ast$ range of the MCs.} %in units of $M_\odot$/sr
%
%Mdust peaks at 6e7 (5-7e7) Mo/sr or 240 Mo/pc2 (LMC); 8e7 (7-10e7) Mo/sr or 220 Mo/pc2 (SMC)
%Mstar peaks at 5e10 (3.5-6.5) Mo/sr or 2.0e5 Mo/pc2 (LMC); 4.5e10 (3-6) Mo/sr or 1.3e5 Mo/pc2 (SMC)
% at a distance of 50 kpc, then 1 sr equiv 250 kpc^2.  lmc has a radius of ~2.15 kpc (check this), so a surface area of ~14.47 kpc^2
% note that $M_\odot/pc^2$ is about 1/250000 (10**5.4) lower than $M_\odot/sr$; for the smc, this is 1/360000 (10**5.6)
The resulting stellar mass ($M_\ast$) maps have typical values of approximately 
%$4.0\times10^{10}M_\odot/$sr and $3.6\times10^{10}M_\odot/$sr in the LMC and SMC, respectively. 
$120~M_\odot/\mathrm{pc}^2$ and $70~M_\odot/\mathrm{pc}^2$ in the LMC and SMC, respectively. 
(The dust masses in Section~\ref{sec:Mdustdesc} are lower, having typical values of approximately 
%$6.0\times10^7M_\odot/$sr and $8.0\times10^7M_\odot/$sr in the LMC and SMC, respectively.
$0.07M_\odot/\mathrm{pc}^2$ and $0.02M_\odot/\mathrm{pc}^2$ in the LMC and SMC, respectively.) 
By integrating over the stellar mass (surface density) maps, we obtain total masses of 
$2.0\times10^9~M_\odot$ for the LMC and $3.1\times10^8~M_\odot$ for the SMC, consistent with Harris \& Zaritsky (2004, 2009). %maybe also cite van der Marel+ 2002; Stanimirovic+ 2004
%XXXOLD: total stellar masses 1.228e9 for LMC and 1.601e8 for SMC
% new total stellar masses: 1.726e9 for LMC and 3.966e8 for SMC (maybe 3.4-3.9e8 is more realistic)
%
%[I should estimate uncertainties of these global values.]
% wtf?  now I'm getting 2.276e9 and 3.383e8 for the total stellar masses. this is from summing the pixels; when I use the total fluxes, I obtain these masses: 1.744e9 and 2.754e8. 
%
% new: I'm revising the total masses, raising the LMC by 15% to 1.99e9Msun, and lowering the SMC's by 20% to 3.05e8Msun.

\subsubsection{Star Formation Rate}\label{sec:SFRdesc}

We now use the combination $H\alpha+24~\mu$m as a star formation rate indicator, 
in order to account for both obscured and unobscured star formation (e.g., Calzetti et al.\ 2007; Kennicutt et al.\ 2009; Leroy et al.\ 2012). 

We use $H\alpha$ data from the SHASSA survey (Gaustad et al.\ 2001). 
We did not attempt to remove contamination by [N\textsc{II}] emission lines, which could contribute more than 20\% relative to $H\alpha$ for some objects (Gaustad et al.\ 2001), but is of order 10\% or less for faint dwarf and late-type galaxies like the MCs (Kennicutt et al.\ 2008) and is not expected to significantly affect the resulting SFRs. 

%[also \textbf{describe MCELS} (Smith et al.\ 1999) data for SMC]. 
The SHASSA images are not very deep, and their beam size is larger than that of SPIRE 500~$\mu$m ($\sim76''$). %($\sim45''$). 
SHASSA uses a $H\alpha$ filter centered at 6563\AA$\,$ with a 32\AA$\,$ bandwidth.  
The LMC and SMC are in fields 013 and 510, respectively, and for the analysis presented here, the SHASSA continuum-subtracted map has been used. 
%we first convolve the $H\alpha$ images to a common (500$\mu$m) resolution, as was done with the IR images in Section~\ref{sec:data}, 
Since the SHASSA images are at poorer resolution than that of SPIRE 500~$\mu$m, we 
fit a Gaussian PSF to the continuum images, constructed a convolution kernel, and convolved the 24~$\mu$m images to this resolution, 
using the code designed by Gordon et al.\ (2008).  A common resolution is necessary to combine $H\alpha$ and $24~\mu$m luminosities in the SFR calibration. 
The SHASSA images are in units of deci-Rayleighs, which are converted to our flux density units 
%($1\mathrm{R}=4.3487/4\pi~\mathrm{MJy}/\mathrm{sr}$ at $\lambda=6563$\AA). 
($1\mathrm{R}=2.409\times10^{-7}{\rm erg}/{\rm s}/{\rm cm}^2/{\rm sr}$ at $\lambda=6563$\AA). 

% from Lawton10: The Hα images are from the southern sky wide-angle imaging survey known as the SHASSA (see Gaustad et al. 2001). The SHASSA survey uses a Canon camera with a 52 mm focal length lens and an Hα filter centered at 6563 Å with a 32 Å bandwidth. The LMC and SMC images are each comprised of five separate 20 m exposures. The LMC and SMC are in fields 013 and 010 [sic] in Gaustad et al. (2001), respectively. SHASSA images have	a similar resolution (∼45′′ ) to that of	the	MIPS 160 μm images (40′′ ).
% from Bernard08: The survey has been carried out with a narrow-band imaging system consisting of an Hα filter of 3.2 nm bandwidth and a dual-band notch filter that excludes Hα but transmits a 6.1 nm band of continuum radiation on either side of Hα, centered at 644 and 677 nm. For the analysis presented here, the continuum-subtracted map has been used. The survey angular resolution is 0.8′, which corresponds to a linear resolution of 11.6 pc. The sensitivity reaches 2 Rayleighs (1 Rayleigh = 106 photons cm−2 s−1 corresponding to 2.41 × 10−7 erg s−1 cm−2) and it is mainly limited by geocoronal contamination.

%[We could briefly compare SHASSA vs MCELS $H\alpha$ for the SMC, as a resolution test.] 

We compute SFRs of the MCs on a pixel-by-pixel basis, using the calibration proposed by Calzetti et al.\ (2010), which is calibrated for normal galaxies and H\,{\sc ii} regions: 
%%which has been shown to be quite accurate, 
%with scatter $<\,0.2$ dex, for galaxies with a wide range of metallicities:
\begin{equation}
  SFR\,(M_\odot/{\rm yr}) = \left\{
  \begin{array}{l}
    \displaystyle C_{{\rm H}\alpha}\,[L({\rm H}\alpha)_\mathrm{obs}+a_1 L(24)] \vspace{0.05cm}\\
    \displaystyle \quad \mathrm{if}\,\, L(24)<4\times 10^{42}\,{\rm erg}/{\rm s}, \vspace{0.2cm}\\
    \displaystyle C_{{\rm H}\alpha}\,[L({\rm H}\alpha)_\mathrm{obs}+a_2 L(24)] \vspace{0.05cm}\\
    \displaystyle \quad \mathrm{if}\,\, 4\times10^{42}\leq L(24) < 5\times10^{43}\,{\rm erg}/{\rm s}, 
  \end{array} \right.
  \label{normalSFR}
\end{equation}
%\begin{eqnarray}
%  SFR\,(M_\odot/{\rm yr})&=&\,C_{{\rm H}\alpha}\,[L({\rm H}\alpha)_\mathrm{obs}+a_1 L(24)] \label{normalSFR} \\
%   &\phantom{=}& \mathrm{if}\,\, L(24)<4\times 10^{42}\,{\rm erg}/{\rm s}, \nonumber\\
%       &=&\,C_{{\rm H}\alpha}\,[L({\rm H}\alpha)_\mathrm{obs}+a_2 L(24)] \label{starburstSFR} \\
%   &\phantom{=}& \mathrm{if}\,\, 4\times10^{42}\leq L(24) < 5\times10^{43}\,{\rm erg}/{\rm s} \nonumber
%\end{eqnarray}
where the luminosities are in units of $\mathrm{erg}/\mathrm{s}$, 
and $C_{{\rm H}\alpha}=5.45\times10^{-42}(M_\odot/{\rm yr})/(\mathrm{erg}/\mathrm{s})$. 
The coefficients $a_1=0.020$, and $a_2=0.031$, are tuned to recover the dust-obscured SFR, as opposed to emission due to transiently heated dust grains (Calzetti et al.\ 2007). 
%The first of these equations (\ref{normalSFR}) is calibrated for normal galaxies 
%(Kennicutt et al.\ 2009), while the latter (\ref{starburstSFR}) is calibrated for HII 
%regions and starbursts (Calzetti et al.\ 2007).  
This calibration has a scatter of less than 0.2 dex, and it assumes a Kroupa %(2001) 
IMF (converting from a Kroupa to Salpeter implies SFRs larger by a factor of $\sim1.5$). %(but should probably convert $M_\ast$ \& SFR to same IMF?). 
%see also SFR discussion in Leroy et al.\ (2012). 
% Alberto and Katie (student) are doing something with star formation, incl possibily Ha+24um
Note that our SFR maps may be inaccurate in regions with diffuse `IR cirrus' dust emission, 
were the dust may be only partially heated by young stars (see Kennicutt et al.\ 2009). 

Our approach differs from the one recently proposed by Leroy et al.\ (2012), which applies detailed dust models to the far-IR SED to account for emission due to the `cirrus' radiation field, rather than simply using the $H\alpha$ and 24~$\mu$m luminosities.  In addition, their star formation surface density is averaged over larger areas ($\sim{\rm kpc}$ resolution) than our analysis of the Magellanic Clouds.  
The 24~$\mu$m contribution can result in up to 0.2 dex uncertainty in the total $H\alpha+24~\mu$m SFR; however, in both the LMC and SMC, we find that the 24~$\mu$m contribution is rather small: 
in Eqn.~\ref{normalSFR}, 80\% and 92\% of the pixels in the LMC and SMC, respectively, have $L(24~\mu{\rm m})<4\times 10^{42}\,{\rm erg}/{\rm s}$. 
% 79.50% in the LMC (1.54672e6/1.94555e6 pixels), and 92.24% in SMC (270056/292780 pixels).

The integrated properties of the LMC and SMC are summarized in Table~\ref{tab:props}.

\begin{table}
 \caption{Integrated Properties of the Magellanic Clouds}
 \centering
 \begin{tabular}{ l | c c c c c }
   \hline
    & $\mathrm{log}~L_\mathrm{dust}$ & $\mathrm{log}~M_\mathrm{dust}$ & $\mathrm{log}~M_\ast$ & $\mathrm{log}~SFR$ \\ %& $M_\mathrm{gas}/M_\mathrm{dust}$ \\
    & (log~$L_\odot$) & (log~$M_\odot$) & (log~$M_\odot$) & (log~$M_\odot~\mathrm{yr}^{-1}$) \\ %& \\
   \hline
   %dust luminosities: 2.3e8 & 2.5e7
   %dust masses: 1.07e6 & 1.09e5
   %stellar masses: 2.0e9 & 3.1e8
   %SFRs: 0.38 & 0.024
   LMC & $8.4\pm0.15$ & $6.0\pm0.1$ & $9.3\pm0.1$ & $-0.4\pm0.2$ \\ %& $340\pm40$ \\
   SMC & $7.4\pm0.15$ & $5.0\pm0.1$ & $8.5\pm0.1$ & $-1.6\pm0.2$ \\ %& $900\pm150$ \\
   \hline
  \end{tabular}
 \begin{list}{}{}
    \setlength{\itemsep}{0pt}%{1pt}
    \item The quantities are computed by integrating the resolved maps described in this paper.  For details, see Sections~\ref{sec:Ldustdesc}, \ref{sec:Mdustdesc}, \ref{sec:Mstardesc}, and \ref{sec:SFRdesc}, about the dust luminosity, dust mass, stellar mass, and SFR, respectively.
          %footnote for gas/dust.
 \end{list}
 \label{tab:props}
\end{table}

%%%------------------------------------------------------------------------------%%%

\section{Distribution of Dust Luminosity and Mass}\label{sec:results}

%In this section, we describe and analyze the dust properties of the Magellanic Clouds using their dust SEDs. 
%We first discuss measures of dust luminosities in Section~\ref{sec:Ldust}, inferred from the mid-IR to far-IR SEDs. 
%In Sections~\ref{sec:Tdust} and \ref{sec:Mdust}, we discuss the dust temperatures and masses, which are estimated from modified blackbody fitting and are taken from Gordon et al.\ (in prep.). 

%\subsection{Dust Luminosity}\label{sec:Ldust}

%\subsubsection{Description}
% moved to above section and Appendix A
%\subsubsection{Results}

\begin{figure*}
 \includegraphics[width=0.497\hsize]{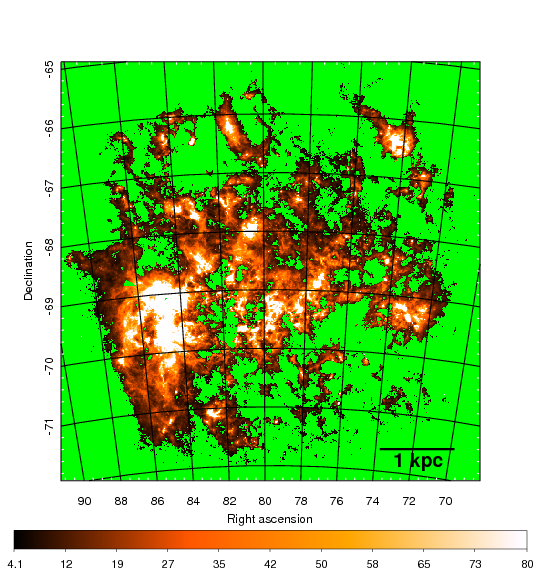}%{LMC_Ldust_Lsun_Karlnanbg_axes.png} %{LMC_Ldust_Lsun_Karlnanbg.png} 
 \includegraphics[width=0.497\hsize]{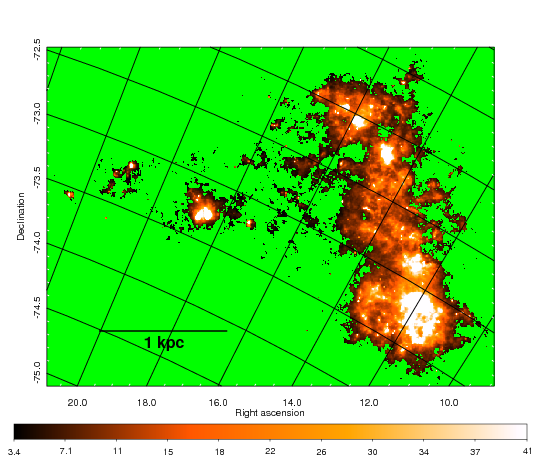}%{SMC_Ldust_Lsun_Karlnanbg_axes.png} 
 \caption{LMC (left) and SMC (right) $L_\mathrm{dust}$ maps, estimated by summing 5.8-500~$\mu$m photometry (see text for details), in units of $L_\odot/{\rm pc}^2$, as indicated by the color bars below the panels.  
  The right ascension and declination coordinates are in units of degrees; the coordinates are omitted from subsequent figures.  
  Green regions are masked, because they have no or noisy far-IR data, or because the modified blackbody fits are poor.}
 \label{fig:Ldust}
\end{figure*}
%% I also have a plot of Mdust/LTIR: LMC_MdustLTIR_ratio.png,  but I noticed that the inverted ratio looks a lot like the Tdust map, which is probably not surprising, since by our definition, L500/Mdust is proportional to the Planck function B(Tdust).
%% Chad: That [Mdust/LTIR] seems to do a pretty good job of showing where the HII regions are.  It might be of interest (not for the paper, though - this is just the kind of thing to do if you're curious) to compare that to an Halpha map.

We begin by analyzing the distribution of dust luminosity.  
As stated in Section~\ref{sec:Ldustdesc}, these utilize \textit{Herschel} data, integrating the luminosity at $5.8~\mu{\rm m}\leq\lambda\leq500~\mu{\rm m}$. 

In Figure~\ref{fig:Ldust}, we show the dust luminosity maps of the LMC and SMC. 
Here and throughout the paper, we apply a far-IR signal-to-noise cut, following Gordon et al.\ (in prep.). 
%Within both MCs, the dust luminosity is typically in the range of $50-300~L_\odot$ per square pixel, 
%with the distributions peaking at $L_\mathrm{dust}\sim100~L_\odot$ per square pixel; 
The dust luminosity is typically in the range of $4-55~L_\odot/{\rm pc}^2$ within the LMC and $3-30~L_\odot/{\rm pc}^2$ within the SMC, 
with the distributions broadly peaking at $L_\mathrm{dust}\sim7-11$ and $5-8~L_\odot/{\rm pc}^2$, respectively. 
The $L_\mathrm{dust}$ distributions of the MCs have ``tails" at the bright end, as can be seen in the bright regions in the figure, and the distributions are not affected by the signal/noise threshold used to make the masks, as pixels with noisy data and poor S/N have been masked out. 
In Appendix~\ref{app:LTIR}, we compare this dust luminosity to the total IR luminosity inferred from shorter wavelengths ($\lambda\leq160~\mu$m), and show that they have only small differences. 

%\subsection{Dust Temperature}\label{sec:Tdust}

%\subsubsection{Description}
%\subsubsection{Results}

%\subsection{Dust Mass}\label{sec:Mdust}

%\subsubsection{Description}
%\subsubsection{Results}

%\begin{figure*}
% \includegraphics[width=0.497\hsize]{LMC_Mdust500_fixed.png}
% \includegraphics[width=0.497\hsize]{SMC_Mdust500_fixed.png}
% \caption{LMC and SMC dust mass distributions, inferred from modified blackbody fit to FIR and submm fluxes.}
% \label{fig:Mdust}
%\end{figure*}
\begin{figure*}
 \includegraphics[width=0.497\hsize]{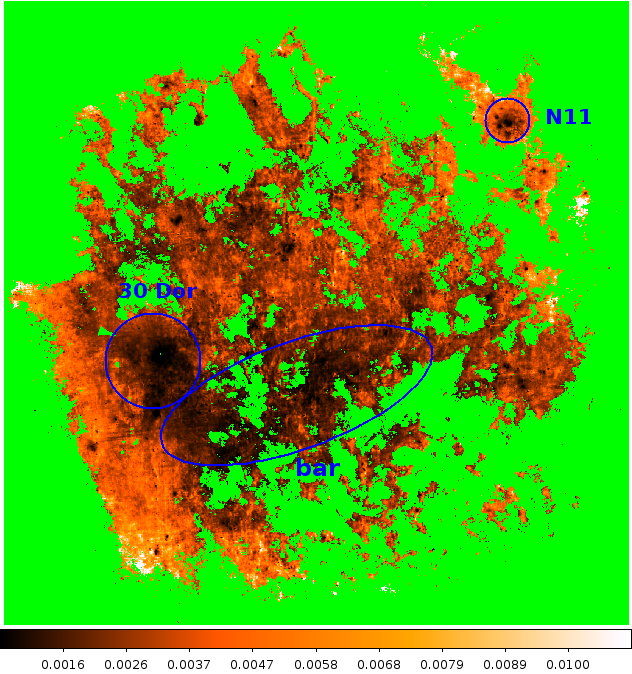}%{LMC_MdustLdust_fig2a_v2.png}%fig2apng.png %{LMC_KarlsMdustLdust_ratio_MLsun2.png} 
 \includegraphics[width=0.497\hsize]{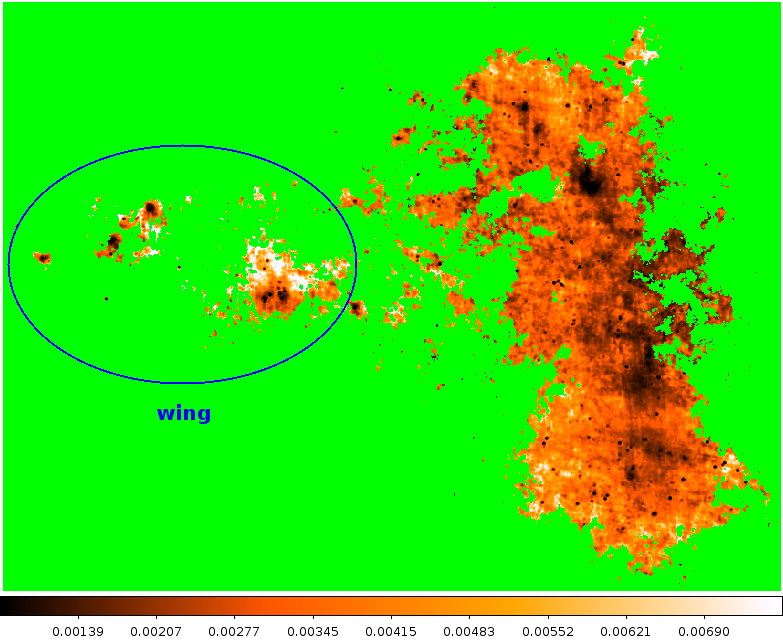}%{SMC_MdustLdust_fig2b_v2.png}%fig2bpng.png %{SMC_KarlsMdustLdust_ratio_MLsun2.png} 
 \caption{Maps of dust mass/luminosity ratio of LMC (left) and SMC (right), in units of $M_\odot/L_\odot$. %($M_\odot$/$10^8$)/(erg/s/$10^{44}$)/sr.
  A few selected regions that are discussed in the text are labeled.} 
  %Green regions are masked, because they have no or noisy far-IR data, or because the modified blackbody fits are poor.
 \label{fig:MLdust}
\end{figure*}

%we now show/discuss dust mass distributions [see Karl's paper], 
%in units of $M_\odot/\mathrm{sr}$ (\textbf{right?}). %or Msun/pixel? 
%\textbf{[Karl's $T_\mathrm{dust}$ \& $M_\mathrm{dust}$ maps are similar to ours.  let's use his, but mention the small differences btw them.]}

% my current M/Ldust maps use my dust masses; let's look at the ratio using Karl's dust masses. 
We now can compare the dust mass to dust luminosity, using the 
$M_\mathrm{dust}/L_\mathrm{dust}$ ratio, which is shown for the LMC and SMC in Figure~\ref{fig:MLdust}. 
$M_\mathrm{dust}/L_\mathrm{dust}$ typically ranges from 0.001 to $0.005~M_\odot/L_\odot$ 
in both MCs, with a mean of $0.0026~M_\odot/L_\odot$. 

The $M_{\rm dust}/L_{\rm dust}$ ratio has an inverted distribution with respect to that of $T_{\rm dust}$, such that it tends to be low in regions where $T_{\rm dust}$ is high, such as within and near the bar of the LMC.  
%This is as expected, because dust luminosity scales roughly as 
%$L_\mathrm{dust}\propto T_\mathrm{dust}^6$ for a dust emissivity exponent of $\beta=2$.  
%and is simply explained by 
%the fact that dust heating (by various stellar populations and the diffuse radiation field) and emission from dust grains tend to occur in the same regions.

%[Maybe show plots of $L_{\rm d}/M_{\rm d}$ vs $T_{\rm d}$ and $L_{\rm d}$ vs $T_{\rm d}$?  
%The former has a strong correlation, but the latter has considerable scatter especially at faint $L$. implications for relation btw shape of SED \& dust temp.] 
%relevant for dust emission models (e.g., Lacey+ 2008, Somerville+ 2012)? 
\begin{figure}
 \includegraphics[width=\hsize]{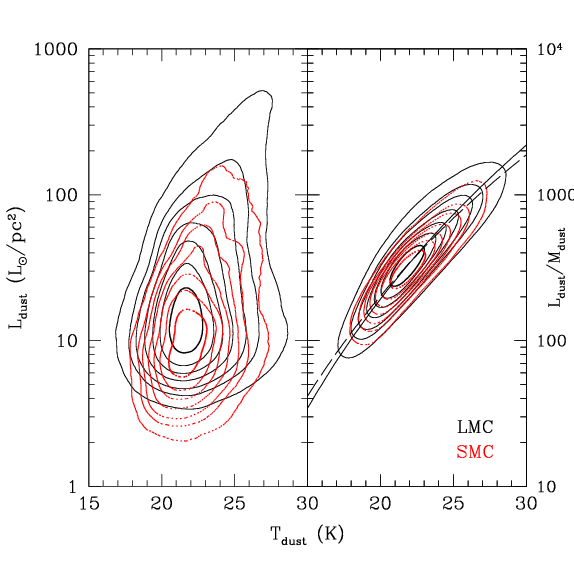}%{LMCandSMC_LdustvsTdust_contour_2panels.png}
 \caption{LMC (black solid contours) and SMC (red dotted contours) distributions of $L_\mathrm{dust}$ (left, in $L_\odot/{\rm pc}^2$ units) and $L_\mathrm{dust}/M_\mathrm{dust}$ (right, $L_\odot/M_\odot$) versus $T_\mathrm{dust}$.  
          In the right panel, the solid and dashed lines through the contours show $L_\mathrm{dust}/M_\mathrm{dust}\propto T_\mathrm{dust}^{4+\beta}$ for $\beta=2$ and 1.5, respectively (see text for details).}
 \label{fig:LdustTdust}
\end{figure}
We show this explicitly in Figure~\ref{fig:LdustTdust}, plotting the $L_\mathrm{dust}$ and $T_\mathrm{dust}$ distributions of pixels in the LMC and SMC. 
Approximately 1,000,000 and 110,000 pixels were used for the LMC and SMC, respectively.  
We also plot $L_{\rm dust}/M_{\rm dust}$ versus $T_{\rm dust}$ in the right panel of Figure~\ref{fig:LdustTdust}. 
It clearly shows that the correlation between them is much stronger than the 
$L_\mathrm{dust}$--$T_\mathrm{dust}$ correlation: 
% spearman rank: 0.26 \& 0.22 for SMC \& LMC for L; 0.95 \& 0.92 for ratio
the former has Spearman rank correlation coefficients of $r_s=0.95$ and 0.92 for the SMC and LMC, respectively, while the latter has much weaker values of 0.26 and 0.22.\footnote{The Spearman rank correlation coefficient may have a value between $-1$ and 1.  A positive (negative) value indicates an (anti)correlation, and a value of 0 indicates no correlation.} 
%Although there is a connection between $L_\mathrm{dust}$ (i.e., dust emission) and the shape of the dust SED (i.e., dust heating), this clearly indicates that it is rather the dust luminosity/mass ratio that is the more closely related quantity.  
It is simply due to the fact that $L_\mathrm{dust}\propto M_\mathrm{dust}\times T_{\rm dust}^{4+\beta}$ 
(as shown by the solid lines in Fig.~\ref{fig:LdustTdust}), while the $L_\mathrm{dust}$--$T_{\rm dust}$ correlation involves
the dust column distribution. %such that $L_\mathrm{dust}$ can be low in regions with high $T_{\rm dust}$ provided that the dust column is low.
For example, there are some regions, such as the southern edge of the LMC bar, that have warm temperatures but 
not extremely high luminosities; %; however, they simply have large luminosities per unit mass. 
by normalizing $L_{\rm dust}$ with $M_{\rm dust}$, one effectively removes the dependence on dust column. 
Furthermore, note that the slope of the SMC's correlation in the right panel is slightly shallower than the LMC's, which implies a slightly lower emissivity index $\beta$ in the SMC. 
%In addition, from equation (\ref{Mdust}), we see that $B_{160\mu{\rm m}}(T_{\rm dust})\propto L_{160\mu{\rm m}}/M_{\rm dust}$, such that the correlation between $T_\mathrm{dust}$ and $L_\mathrm{dust}/M_\mathrm{dust}$ is not unexpected; normalizing by dust mass effectively removes the dependence on dust column density. 
%This is also related to the issue that dust mass determinations depend on the spatial scales probed: with high resolution, we can probe cold components shielded by the emission from warmer regions (Galliano et al.\ 2011).

Because of the expected spatial variation of dust column, one would not 
expect a one-to-one correspondence between $L_{\rm dust}$ and $T_{\rm dust}$. 
%This explains the apparent bimodal $L_{\rm dust}$--$T_{\rm dust}$ 
This explains the tail towards bright $L_\mathrm{dust}$ in the 
distribution in the left panel of Figure~\ref{fig:LdustTdust}. 
This can be understood such that bright regions 
($L_{\rm dust}>50L_\odot/{\rm pc}^2$) are mostly coincident with dusty 
star-forming regions, because $L_\mathrm{dust}$ is an indicator of obscured 
star formation (see Sec.~\ref{sec:SFRdesc}), while the fainter regions 
generally probe the diffuse ISM.
%Lastly, one could interpret the left panel of Figure~\ref{fig:LdustTdust} as a 
%bimodal distribution of dust luminosity and temperature: bright regions ($L_{\rm dust}>700L_\odot/{\rm pixel}$) are for the most part coincident with dusty star-forming regions, because $L_\mathrm{dust}$ is an indicator of obscured star formation (see Sec.~\ref{sec:SFRdesc}), while the fainter regions generally probe the diffuse ISM.

\section{Dust vs Stellar Distribution}\label{sec:dusttostellar}

In this section, we show and analyze the spatial distribution of the LMC and SMC's dust/stellar luminosity and mass ratios, using the quantities defined in Section~\ref{sec:description}, inferred from the resolved SEDs. 
%[\textbf{usefulness of the ratio} (S11).]
Galaxy SEDs, and dust/stellar ratios in particular, are related to other properties indicative of a galaxy's evolution, such as metallicity, stellar mass, starlight intensity, and morphology (e.g., Fontanot et al.\ 2009; da Cunha et al.\ 2010). %Groves et al. 2008, Jonsson et al. 2010 
The dust/stellar luminosity ratio can be physically interpreted as the amount of emission being reprocessed by dust (both small and large grains) relative to the unobscured emission from 
(young and old) stars, 
while the dust/stellar mass ratio can be interpreted as the relative amounts of stellar mass growth and dust production.  As shown by S11, dust/stellar luminosity and mass ratios are not equivalent, and have substantial scatter between them for dwarf and late-type galaxies. 

\subsection{Spatial Distribution of Dust/Stellar Luminosity}\label{sec:Lratio} %Dust/3.6$\mu$m

%\subsubsection{Stellar Luminosity Calibration} 
%\subsubsection{Results: Spatial Distribution of Dust/Stellar Luminosity}

\begin{figure*}
 \includegraphics[width=0.497\hsize]{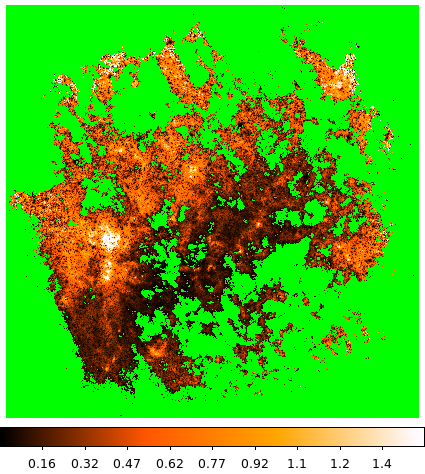}%{LMC_LdustLstarbolconst_ratio_Karlnanbg.png} %{LMC_LdustLstar_ratio_new_Karlnanbg.png} %{LMC_LdustLstar_ratio.png}
 \includegraphics[width=0.497\hsize]{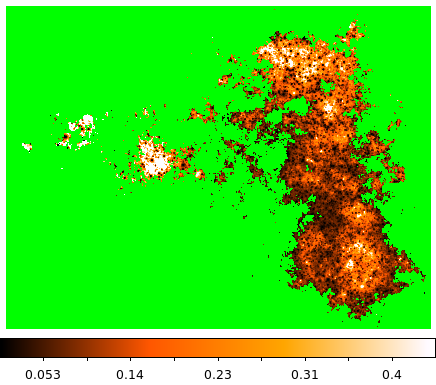}%{SMC_LdustLstarbolconst_ratio_Karlnanbg.png} %{SMC_LdustLstar_ratio_new_Karlnanbg.png} %{SMC_LdustLstar_ratio.png}
 \caption{LMC (left) and SMC (right) dust/stellar luminosity (unitless) ratio maps.  The dust and stellar luminosities are described in Sections~\ref{sec:Ldustdesc} and \ref{sec:Lstardesc}.} %see text for details. %, where the $L_\mathrm{dust}$ is estimated as in Section~\ref{sec:Ldust} and $L_{3.6\mu{\rm m}}$ is used as a proxy for $L_\ast$.}
 \label{fig:LdustLstar}
\end{figure*}

We present maps of the $L_\mathrm{dust}/L_\ast$ luminosity ratios of the LMC and SMC in Figure~\ref{fig:LdustLstar}. 
%We first present maps of the luminosity ratio $L_\mathrm{dust}/L_{3.6\mu{\rm m}}$ (where $L_\mathrm{dust}$ is estimated from Eqn.~\ref{eq:fdust}), 
%which we will interpret as an imperfect proxy for $L_\mathrm{dust}/L_\ast$.  
%These luminosity ratios are shown for the LMC and SMC in Figure~\ref{fig:LdustLstar}.  
%
The dust/stellar luminosity ratios exhibit substantial spatial variation within both galaxies. 

The values of $L_\mathrm{dust}/L_\ast$ typically range from %0-25 and 0-50  
0.01-0.2 and 0.01-0.6 throughout the SMC and LMC, respectively. %peaking at a value of $\approx10$.
%0-80 throughout the SMC (peaking at $\sim200$) and 0-150 in the LMC (peaking at $\sim400$). 
These are comparable to the global values estimated by S11, who obtained 
$L_\mathrm{dust}/L_\ast\sim0.1-0.5$ for nearby metal-poor ($Z<0.4 Z_\odot$) dwarf galaxies. %, because their $L_\ast$ were estimated by integrating the stellar SED, rather than using only 3.6$\mu$m luminosity.

%\textbf{interpretation of regions with high vs low $L_\mathrm{d}/L_{3.6}$?}
Many regions of the MCs that have bright dust luminosities or large dust mass, such as 30~Dor and the SMC ``bar", not surprisingly also 
have high $L_\mathrm{dust}/L_\ast$ ratios. 
%\textbf{[explain why high ratio in 30 Dor \& SMC ``bar".]} %Jacco: I think we should try to make sense of why the Ldust/Lstar ratio is so high in 30 Dor, in the SMc Wing, but not in the South-East of the LMC and why it is more concentrated in the SMC "bar". It is largely related o hte fact that the higher dust densities are obtained in dense clouds whereas much of the stellar light is distributed much more uniformly, especially the infrared light of old and intermediate-age post-main-sequence stars.
In contrast, some regions have particularly low ratios, such as the LMC bar (see Figs.~\ref{fig:Ldust} and \ref{fig:LdustLstar}), which has a relatively large stellar flux density (Harris \& Zaritsky 2009). 
Here, dark spots can be seen, and are attributed to regions with stars and star clusters (which are resolved in spite of the fact that the $L_\ast$ maps are degraded to SPIRE $500~\mu$m resolution), lacking significant far-IR emission from dust.

%[issue of continuous dust dist vs discrete stars.]
It is important to keep in mind that stars are discrete objects, while dust grains have a more continuous spatial distribution.  In addition, most dust is heated by starlight in the diffuse radiation field, while dust near compact star-forming regions are heated more intensely (e.g., Aniano et al.\ 2012; Foyle et al.\ 2012). 
It is beyond the scope of this paper to investigate emission from individual stars (see e.g. Harris \& Zaritsky 2004, 2009), though such an approach is complementary to ours. 
In any case, our dust/stellar ratios are not very sensitive to the smoothing scale, and in Section~\ref{sec:sfr} we examine star formation activity using smoother $H\alpha$ luminosity maps.

\begin{figure}
 \includegraphics[width=\hsize]{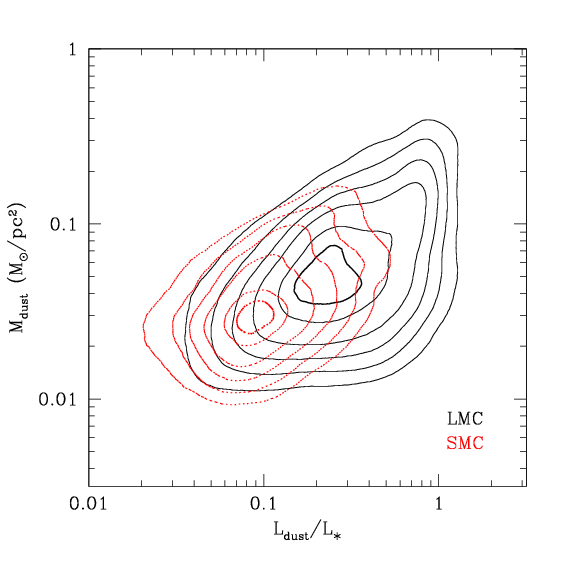}%{LMCandSMC_LratiovsMdust_contour.png}
 \caption{LMC (black solid contours) and SMC (red dotted contours) distributions of $L_\mathrm{dust}/L_\ast$ versus $M_\mathrm{dust}$.}
 \label{fig:LratiovsM}
\end{figure}

One might expect that the dust luminosity is roughly proportional to the total dust mass multiplied by the stellar luminosity (i.e., $L_{\rm dust} \propto L_\star\times M_{\rm dust}$), 
based on the energy balance of the dust between the absorption of starlight 
(which is proportional to $L_\star\times M_{\rm dust}$) and the emission 
of IR photons (which is essentially $L_{\rm dust}$).  
To first order, the spatial variation of $L_{\rm dust}/L_\star$ roughly reflects 
the spatial variation of $M_{\rm dust}$, as shown by the weak correlation in Figure~\ref{fig:LratiovsM} (cf. Fig.~4 of S11, for whole galaxies).  However, 
this is complicated by the dependence 
on the radiative environment, due to the fact 
that the heating of the dust by starlight is determined by the intensity
and spectral distribution of the starlight (e.g., depending on the dust's proximity to 
star-forming regions or whether it is heated by old stars or the diffuse ISRF) and 
by the intrinsic absorption and emission properties of the dust (which is 
determined by its composition and size distribution). 
It is important to keep in mind that the stellar emission is due to a variety of stellar populations with different ages, and the dust emission is due to dust grains of different compositions and sizes. 
%The radiative environment matters: the dust/stellar balance in a particular region is affected by its proximity to star-forming regions and heating due to old stars or the diffuse ISRF. 

%\textbf{show contour plot of $L_\mathrm{dust}/L_\ast$ vs $M_\mathrm{dust}$.}

In addition, as we have defined $L_\ast$, it is more sensitive to light due to older rather than younger stellar populations.  
The highest $L_\mathrm{dust}/L_\ast$ ratios in Figure~\ref{fig:LdustLstar} often occur near major H\,{\sc ii} region complexes, and may be caused by dust directly heated by OB stars and generally hotter than elsewhere, but may also reflect a poor recovery of the stellar luminosity of the luminous OB star clusters exciting these regions (i.e., underestimating $L_\ast$) by using only 3.6 and 4.5~$\mu$m emission (see Sec.~\ref{sec:Lstardesc}). 
% Frank: Somewhere in Section 5.1. I would explicitly remark that the highest Ld/L* ratios almost invariably occur towards major HII region complexes, and may be caused by dust directly heated by OB stars and generally hotter than elsewhere, but almost certainly also reflect relatively poor recovery of the actual stellar luminosity of the luminous OB star clusters exciting these regions (underestimating L*) using only 3.5/4.5mu NIR emission.
% I am mentioning this because in Figure 4, one recognizes immediately and very clearly the HII region distribution in the LMC and the main body of the SMC.

\subsection{Spatial Distribution of Dust/Stellar Mass}\label{sec:Mratio}

%\subsubsection{Stellar Mass Calibration}

%\subsubsection{Results: Spatial Distribution of Dust/Stellar Mass}

\begin{figure*}
 \includegraphics[width=0.497\hsize]{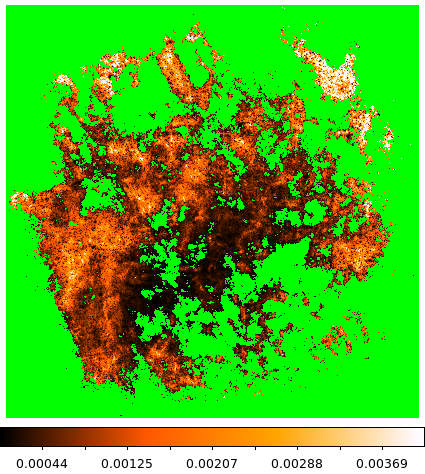}%{LMC_KarlMdustMstar_new.png} %{LMC_MdustMstar.png}
 \includegraphics[width=0.497\hsize]{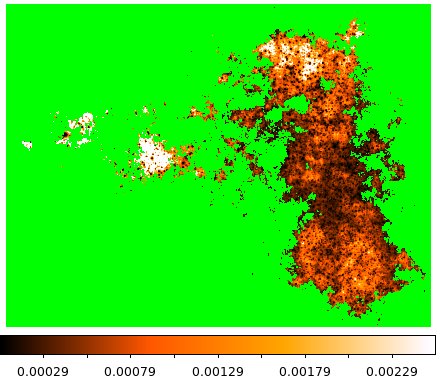}%{SMC_KarlMdustMstar_new.png} %{SMC_MdustMstar.png}
 \caption{LMC (left) and SMC (right) dust/stellar mass ratio maps, where $M_\mathrm{dust}$ is estimated as in Section~\ref{sec:Mdustdesc} and 3.6~$\mu$m and 4.5~$\mu$m are used as proxy for $M_\ast$ (see text for details). 
 %\textbf{[NB: if we use Karl's $M_\mathrm{dust}$, these will look much different than Fig.~\ref{fig:LdustLstar}.]}
         }
 \label{fig:MdustMstar}
\end{figure*}

We now show maps of the mass ratio $M_\mathrm{dust}/M_\ast$ in Figure~\ref{fig:MdustMstar}. 
% where $M_\ast$ is estimated from 3.6 \& 4.5$\mu$m luminosity. 
These are qualitatively similar to the maps of dust/stellar luminosity (Fig.~\ref{fig:LdustLstar}), implying that the mass and luminosity distributions are similar. 
%though some regions that have exceptionally bright dust luminosities have only slightly large $M_\mathrm{dust}/M_\ast$. %<-- is this true? 
Note that 30~Dor %, the bright star-forming region in the LMC, 
has a $M_\mathrm{dust}/M_\ast$ that is larger than average, but its $L_\mathrm{dust}/L_\ast$ is particularly large. 

%most regions in the MCs have $0.0001<M_\mathrm{dust}/M_\ast<0.005$ \textbf{(check this)}.
% LMC: peaks at 0.0002 but many regions up to 0.004
% SMC: peaks at 0.0003 but some regions up to 0.002
Most regions in the MCs have $M_\mathrm{dust}/M_\ast<0.001$ though some regions in the LMC and SMC have ratios up to approximately 0.004 and 0.002, respectively. 
(The gradient at the northern edge of the LMC is an artifact, due to the different background-subtraction performed by Gordon et al., in prep.) %Karl did a *gradient* background subtraction, and Jean-Philppe mentioned a foreground issue too.
We find that the $M_\mathrm{dust}/M_\ast$ distribution peaks at medians of approximately $3\times10^{-4}$ and $4\times10^{-4}$ for the LMC and SMC, respectively, while $L_\mathrm{dust}/L_\ast$ peaks at values of 0.14 and 0.07. 
This is consistent with S11, who observed that $L_\mathrm{dust}/L_\ast$ is slightly more metallicity-dependent than $M_\mathrm{dust}/M_\ast$ in nearby late-type and dwarf galaxies. 

The integrated dust/stellar mass ratio for the whole LMC and SMC are 
$(M_\mathrm{dust})_\mathrm{tot}/(M_\ast)_\mathrm{tot}=5.4\times10^{-4}$ and $3.6\times10^{-4}$, respectively\footnote{To be clear, note that these are ratios of sums, $(\Sigma M_{\rm dust})/(\Sigma M_\ast)$, which are not equivalent to the average ratios, $\langle M_{\rm dust}/M_\ast\rangle$, of the pixels.}. 
We will discuss these ratios in the context of star formation and gas in the next section. 

We can compare the LMC and SMC's $M_\mathrm{dust}/M_\ast$ to results for other galaxies (Dunne et al.\ 2011; S11; Cortese et al.\ 2012).  
%these studies show that gas-poor and early-type galaxies tend to have low $M_\mathrm{dust}/M_\ast$, while dwarf and late-type galaxies with more gas tend to have higher ratios. also redshift dependence, lack of enviro dependence.
%[disagreement with S11?] 
Most galaxies have $10^{-4}<M_\mathrm{dust}/M_\ast<10^{-2}$, though the precise values depend on the assumptions involved in both mass estimates.  
For example, Cortese et al.\ (2012) estimate their stellar masses from optical rather than near-IR luminosities and colors, and assume a Chabrier IMF, resulting in lower stellar masses and hence higher $M_\mathrm{dust}/M_\ast$.  Nonetheless, the LMC and especially the SMC are on the low end of the dust/stellar mass ratios in these studies: only three ($\approx1/5$th) of the dwarf/irregular galaxies in S11 (Holberg~II, NGC~5408, and NGC~3077) have similarly low ratios, though like the MCs, NGC~3077 is tidally interacting with its neighbors, and a significant amount of dust has been stripped (Walter et al.\ 2011).  On average, the SMC's dust/stellar ratio is nearly a factor of ten lower than that of low-mass dwarfs, while the LMC (which is a one-armed spiral) has a $\approx2\times$ lower ratio compared to similar late-type and Magellanic spirals. 
Dunne et al.\ (2011), S11, and Cortese et al.\ (2012) show that dwarf and late-type galaxies that are not gas-deficient tend to have higher ratios, as do galaxies with higher specific star formation rates; however, no strong dependence on metallicity is observed. 
%The MCs are consistent with these broad trends; for example, 
The MCs are consistent with the dependence on gas content: they have slightly low (by a factor of two) dust/gas mass ratios compared to galaxies with similar metallicities (Draine et al.\ 2007; Herrera-Camus et al.\ 2012) consistent with their relatively low dust/stellar masses.  

%[more discussion?]  Maud wanted to know why some regions have such high ratios (>0.001).

%\subsection{Dust/UV and Optical Luminosity Ratio}
%
%we would have to convolve the images to the poorest resolution, which would hopefully yield blurry dust/stellar maps. 
%%could still be interesting though, and would want to compare to Tdust and other maps from Section~\ref{sec:results}.
%(how hard would it be to convolve the images, assuming simple Gaussian PSF?  need to talk to Karl and Gonzalo.) 
%some of these UV and optical data have not been published, so even blurry images would be better than nothing.  we could also compare these to the 3.6$\mu$m images, to try to determine how realistic an indicator of stellar light it is (e.g., compare HII regions). 
%
%UV shuttle images of both MCs and rocket data for LMC (Smith et al.\ 1987; Gordon et al.\ 1994). 
%% Smith A. M., Cornett R. H., Hill R. S., 1987, ApJ, 320, 609
%% Gordon K. D., Witt A. N., Carruthers G. R., Christensen S. A., Dohne B. C., 1994, ApJ, 432, 641
%B,V,R optical data from Joel Parker and Greg Bothun (not published), and $H\alpha$ images from Rob Kennicutt\footnote{see \texttt{http://dirty.as.arizona.edu/$\sim$kgordon/research/mc/mc.html}}.
%should ask Dennis Zaritsky about format in which they have stellar flux density maps available. 

\section{Star Formation and Dust Heating}\label{sec:sfr}

%\subsection{Star Formation Rate Calibration}

%\subsection{Results: Spatial Distribution of SFR}

\begin{figure*}%{figure}
 \includegraphics[width=0.497\hsize]{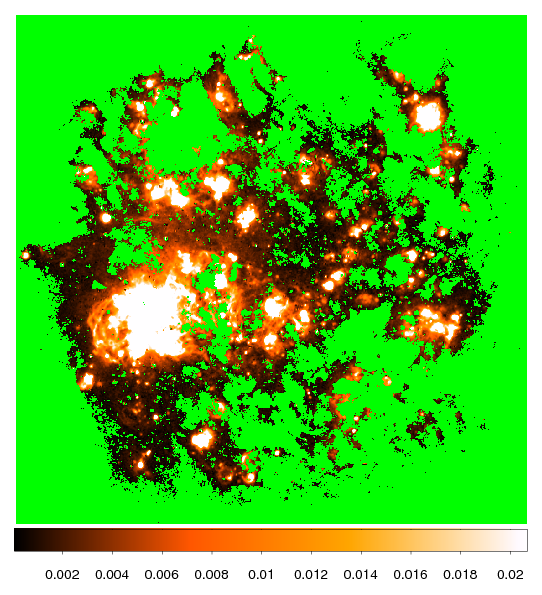} %LMC_SFR_Karlnanbg.png
 \includegraphics[width=0.497\hsize]{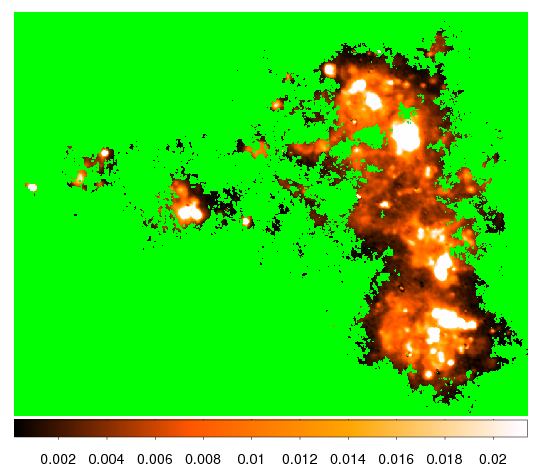} %SMC_SFR_Karlnanbg.png
 \caption{LMC (left) and SMC (right) star formation rate surface density maps, in units of 
          $M_\odot/{\rm yr}/{\rm kpc}^2$.} %previously /sr units rather than /kpc^2
 \label{fig:SFR}
\end{figure*}%{figure}

We present SFR maps of the MCs in Figure~\ref{fig:SFR}.  
They provide an empirical description of the spatial distribution of star formation in the Magellanic Clouds, related to the recent work by Bolatto et al.\ (2011) and Jameson et al.\ (in prep.), and constraints on the galaxies' star formation histories (Harris \& Zaritsky 2004, 2009). 
These SFR maps are also in approximate agreement with the spatial distribution of young stellar objects (YSOs; Carlson et al.\ 2012). 
In Section~\ref{sec:SFRcorrs}, we will show how SFR is related to dust heating and emission. 

%[\textbf{Integrate} to also estimate total SFR of the MCs.] 
% NB: hastrom conserves flux density but not flux
% note that SHASSA_Ha pixels are originally 47.64" and MIPS24um pixels are 2.49", but the SFR maps above have 15.6" pixels? not 14" (500um resolution when I used hastrom)
By integrating over these SFR (surface density) maps, and by accounting for the different 
pixel sizes in order to conserve the total fluxes, we obtain total SFRs: 
$0.38~M_\odot/{\rm yr}$ for the LMC and 
$0.024~M_\odot/{\rm yr}$ for the SMC. 
The MCs' very different SFRs are not surprising considering their stellar masses, which differ by a factor of five; 
moreover, the LMC appears to lie on the `star-forming sequence' of disk-dominated galaxies (e.g., Brinchmann et al.\ 2004; Schiminovich et al.\ 2007). 
Note that, if one were to integrate $L({\rm H}\alpha)$ and $L(24~\mu{\rm m})$ and then apply the SFR calibration, the upper expression of Eqn.~\ref{normalSFR} (calibrated to normal galaxies) yields SFRs approximately $30\%$ lower, 
while the lower expression %Eqn.~\ref{starburstSFR} 
(calibrated to star-forming regions) yields SFRs more similar to the above values. 
% total stellar masses: 1.52e9 for LMC, 1.99e8 for SMC; or 1.2e9 & 1.6e8

%talk to Marta and Alberto about these.  I think Alberto said 0.03 for SMC & Marta said 0.05.
%[compare these values to Marta Sewilo's from YSO counts.] %Sewilo+ (in prep.) for SMC and Whitney B. et al. (2008) for LMC. 
%
%\textbf{SFR discrepancies.} 
These global SFRs are within a factor of two of other estimates in the literature, using different methods, such as YSO counts, stellar population modeling, and far-IR emission. 
Our LMC SFR is larger than that of other estimates (Whitney et al.\ 2008; Harris \& Zaritsky 2009; Lawton et al.\ 2010; Sewi{\l}o et al., in prep.), who obtained 
{$\mathrm{SFR}\approx0.2$-$0.25~M_\odot/{\rm yr}$}. %(see also Jameson \& Bolatto et al., in prep.)
On the other hand, our SMC SFR is lower than that of Bolatto et al.\ (2011) and Wilke et al.\ (2004), who obtained $0.037$ and $0.05~M_\odot/{\rm yr}$, respectively. 
%In addition, our LMC SFR is larger than that obtained from YSO counts (Whitney et al.\ 2008; see also Sewi{\l}o et al., in prep.). 
%%%%%% see Katie's e-mails on 12-13 June. %%%%%
% I'm worried about the LMC SFR: why am I getting a value TWICE that of Alberto & Katie, even though they're also using Halpha+24um calibration?

Finally, note that SFR$(L_{\rm IR})$ (Kennicutt 1998), 
yields much lower values: 0.006 and 0.08 $M_\odot/{\rm yr}$, for the SMC and LMC, consistent with the 
metallicity dependence observed by Dom\'{i}nguez S\'{a}nchez et al.\ (2012).  
These low-metallicity galaxies are dominated by unobscured star formation: 
in `normal' regions (upper expression in Eqn.~\ref{normalSFR}), unobscured star formation accounts for all of the SFR in both galaxies, though in `H\,{\sc ii} regions' (lower expression, with higher $L(24\mu{\rm m})$), obscured star formation accounts for 97\% and 89\% in the LMC and SMC, respectively. 
For a more detailed study of the SFR within the MCs, we refer the reader to Jameson et al.\ (in prep.).

%[Try to compare to \textbf{Harris \& Zaritsky}'s SFH maps?] %also stellar age, SFH
%
The SFRs can be compared to the star formation history maps inferred from optical photometry by Harris \& Zaritsky (2004, 2009) and atomic and molecular gas maps (e.g., Bernard et al.\ 2008; Bolatto et al.\ 2011). 
It is clear that the SFR is coincident with known H\,{\sc ii} regions and regions with high gas column densities,  
and by comparing to previous results in this paper, we see that in some regions the SFR is 
coincident with dust emission (Fig.~\ref{fig:Ldust}) and warm dust (Fig.~\ref{fig:MLdust}) as well. 

\subsection{SFR Correlations}\label{sec:SFRcorrs}

\begin{figure}
 \includegraphics[width=\hsize]{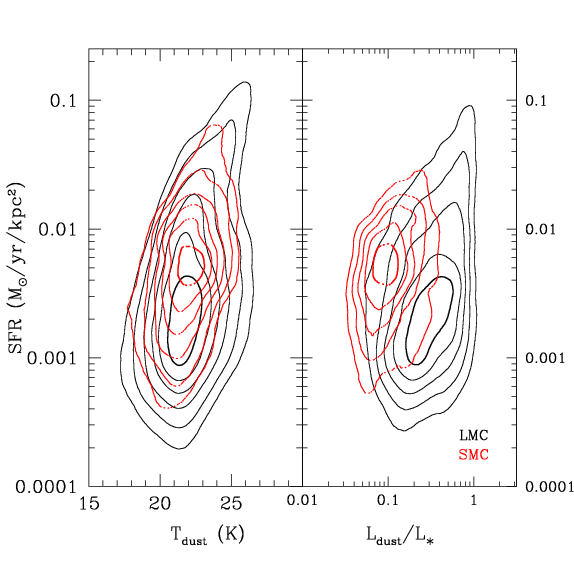}%{LMCandSMC_SFRvsdust_contour_2panels.png}
 \caption{LMC (black solid contours) and SMC (red dotted contours) SFR versus 
          $T_\mathrm{dust}$ (left) and $L_\mathrm{dust}/L_\ast$ (right).  The contours indicate 
          10, 20, 33, 50, and 75\% of the maximum counts, computed from the 
          pixels within the LMC and SMC (Fig.~\ref{fig:SFR}).} %\vspace{0.2cm}}
 \label{fig:SFRcorrs}
\end{figure}

We now examine these correlations directly, by using the SFRs of individual ($47\rlap{.}^{\prime\prime}64$) %($47.64''$) 
pixels. 
SFR versus the dust/stellar luminosity ($L_\mathrm{dust}/L_\ast$) and dust 
temperature is shown in Figure~\ref{fig:SFRcorrs}. 
Because the images have a large number of pixels, %even at this resolution, 
we indicate their distributions with contours. 

Firstly, it is clear that the SMC's median SFR density is a factor of two %3-4  
larger than the LMC's, despite the fact that its total SFR is considerably lower.  
%[Ideas for explanations of this?] 
This is partly due to a projection effect, because the SMC is more inclined %angle of $68^\circ$; 
(Groenewegen et al.\ 2000) than the LMC. 
It may also be due to the different properties of star clusters and H\,{\sc ii} regions in the SMC and LMC (e.g., Lawton et al.\ 2010; Werchan \& Zaritsky 2011), and to a selection effect, as low-SFR pixels are easier to detect in the LMC. 

%Secondly, SFR is weakly correlated with $L_\mathrm{dust}/L_\ast$, and with dust temperature (consistent with Bernard et al.\ 2008; Planck Collaboration 2011), though there is substantial scatter, especially within the LMC. 
Secondly, a correlation between SFR and dust/stellar luminosity or mass has been seen for whole galaxies (da Cunha et al.\ 2010; S11), and we see a similar, though weaker, correlation within the Magellanic Clouds.  A weak correlation with dust temperature is observed as well 
(consistent with Bernard et al.\ 2008; Planck Collaboration 2011). 
The correlation with $T_\mathrm{dust}$ has a Spearman rank of $r_s=0.45$ and 0.36 for the SMC and LMC, respectively, and the correlation with $L_\mathrm{dust}/L_\ast$ has 0.32 and 0.33. 
% spearman rank coefficients $r_s=$ 0.32 \& 0.33 for SMC \& LMC for SFR vs Lratio; 0.45 \& 0.36 for SFR vs Tdust.
It is interesting that there is a pronounced `tail' in these distributions, with stronger correlations at high SFR, $L_\mathrm{dust}/L_\ast$, and $T_\mathrm{dust}$ in the LMC, but 
it has long been known that dust grains may be heated by %young? old?
stars, contributing to far-IR emission (e.g., Calzetti et al.\ 1995; Kennicutt 1998; Walter et al.\ 2007). %Fontanot&Somerville 2011
The substantial scatter around these correlations demonstrates the complex relations between star formation and the wavelength dependence of dust heating and emission. 

% Chad: are there regions with properties (L or M ratio, T) similar to star-forming regions that are *not* forming stars?
There are some regions with high $L_\mathrm{dust}/L_\ast$ that are \textit{not} rapidly forming stars, especially in the LMC, but they generally appear to be in the outskirts of star-forming regions. %explanation of larger spatial extent of dust emission?  
The larger spatial extent of the dust emission may simply be due to the fact that the mean
free path of photons that can heat dust is larger. 
In addition, the southern edge of the LMC's bar, for example, has high $T_\mathrm{dust}$ but less star formation, which is likely an indication of a diffuse `IR cirrus' component of dust emission.

\subsection{Star Formation Efficiencies}\label{sec:SFE}

We measured dust/stellar mass ratios in Section~\ref{sec:Mratio}. 
%[Comment on implications for star/gas, or SF efficiency.] %(star/dust) / (gas/dust)
By inverting the ratios, this implies stellar/dust mass ratios of approximately 1620 and 3350 for the LMC and SMC, respectively. 
These values can be compared to the galaxies' gas/dust ratios, though these have large uncertainties as well, and there is not yet a consensus in the literature on their values. 
We will take the (atomic plus molecular) gas/dust ratios to be 
$(M_\mathrm{gas})_\mathrm{tot}/(M_\mathrm{dust})_\mathrm{tot}\approx340\pm40$ for the LMC (Bernard et al.\ 2008; Galliano et al.\ 2011; see also Meixner et al.\ 2010; Roman-Duval et al.\ 2010) 
and $\approx900\pm150$ for the SMC (Bot et al.\ 2010b; see also Gordon et al.\ 2009; Leroy et al.\ 2011)\footnote{Note that the gas masses used for these gas/dust ratios include a factor of 1.36 to account for the mass of helium (Leroy et al.\ 2011; Galliano et al.\ 2011).}, 
with approximate uncertainties based on the range of the gas/dust estimates in these papers. 
%LMC's GDR. Galliano11: 339, Meixner10: 287: Bernard08: 345
%SMC's GDR. Gordon09: 1200 for tail, Bot10b: 929, Leroy11: 400-500, Bolatto11? 
These yield star/gas mass ratios of $5.5\pm0.8$ and $3.1\pm0.6$ for the LMC and SMC, respectively. 
%with most of the uncertainty due to the gas/dust ratios. 
(In contrast, the MW's star/gas ratio is approximately 10.) %<-- mention metallicity dep. here? 

If we were to define a mass-based `star formation efficiency', $\mathrm{SFE}_m$, as the mass ratio of stars/(stars+gas), which is a common definition on the scale of molecular clouds (Krumholz \& Tan 2007), % also Zamora-Aviles et al. 2012 
then we would have $\mathrm{SFE}_{m,\mathrm{LMC}}\approx0.85\pm0.15$ and $\mathrm{SFE}_{m,\mathrm{SMC}}\approx0.76\pm0.20$. 
% mention Masters et al. 2012: galaxies with high star/dust have a higher strong bar fraction, though the LMC's SFE is high for its stellar mass.
% note that this is different than another definition of SFE=SFR/Mgas, which does not strongly correlate with stellar mass.  [then we want SFR/Mgas=SFR/Mstar*star/gas]  the LMC (but not SMC) appears to have a high SFE by this definition as well (see Huang+12)
% maybe also mention Toribio et al. (2011b).

%[briefly discuss star formation efficiencies.] 
%In Section~\ref{sec:Mratio}, we argued that the LMC and SMC have $M_\ast/(M_\ast+M_\mathrm{gas})\approx0.83$ and 0.79, respectively.  
Furthermore, the LMC has a lower gas mass fraction: $M_\mathrm{gas}/M_\ast\approx18\pm3\%$, versus $32\pm6\%$ for the SMC (though not as low as $10\%$ in the MW).  This is consistent with the fact that more massive galaxies tend to have lower gas fractions (Catinella et al.\ 2010; Masters et al.\ 2012). 
% mention Masters et al. 2012: galaxies with high star/dust have a higher strong bar fraction, though the LMC's SFE is high for its stellar mass.
% note that this is different than another definition of SFE=SFR/Mgas, which does not strongly correlate with stellar mass.  [then we want SFR/Mgas=SFR/Mstar*star/gas]  the LMC (but not SMC) appears to have a high SFE by this definition as well (see Bothwell+09 and Huang+12) & Narayanan12?
% on mass-dep gas fractions, cite Catinella B. et al. (2010, MNRAS, 403, 683) & Masters+12
Another definition of star formation efficiency of galaxies is 
$\mathrm{SFE}_t\equiv\mathrm{SFR}/M_\mathrm{gas}$, whose inverse is the gas consumption time-scale; this SFE only weakly correlates with stellar mass (Bothwell et al.\ 2009; Huang et al.\ 2012; or with dark matter halo mass, in Dav{\'e} et al.\ 2012). 
By this definition, we have $\mathrm{SFE}_{t,\mathrm{LMC}}\approx 10^{-9.0\pm0.1}\mathrm{yr}^{-1}$ 
and $\mathrm{SFE}_{t,\mathrm{SMC}}\approx 10^{-9.6\pm0.1}\mathrm{yr}^{-1}$.  
$\mathrm{SFE}_t$ can be interpreted as the rate at which available gas is converted into stars, 
while $\mathrm{SFE}_m$ quantifies the relative amount of gas converted. 
% SFR/Mstar*(Mstar/Mgas)=10^-8.96 for LMC & 10^-9.69 for SMC
Interestingly, the LMC again appears to have a relatively large $\mathrm{SFE}_t$ for its mass, compared to other low-redshift late-type and dwarf galaxies (Huang et al.\ 2012), while the SMC's is more typical (see also Bolatto et al.\ 2011).  In addition, from its star formation history, the LMC's SFR has been increasing more rapidly with time than the SMC's, on average (Harris \& Zaritsky 2009).

\begin{figure}
 \includegraphics[width=\hsize]{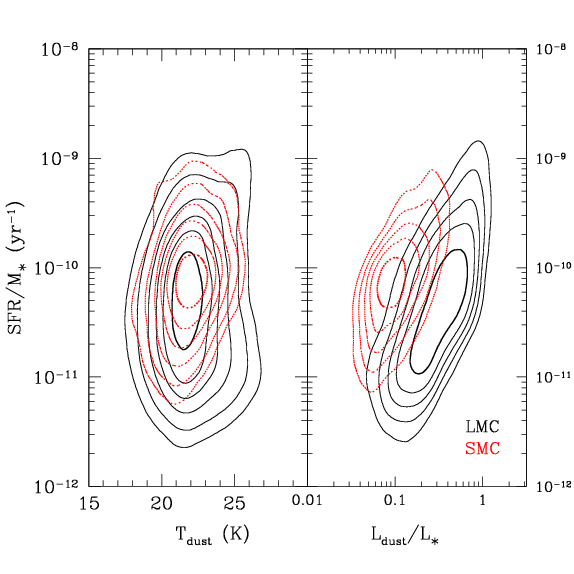}%{LMCandSMC_SSFRvsdust_contour_2panels.png}
 \caption{LMC (black solid contours) and SMC (red dotted contours) SFR$/M_\ast$ versus 
          $T_\mathrm{dust}$ (left) and $L_\mathrm{dust}/L_\ast$ (right), analogous to Figure~\ref{fig:SFRcorrs}.}
          %The contours indicate 10, 20, 33, 50, and 75\% of the maximum counts, 
          %computed from the pixels within the LMC and SMC (Fig.~\ref{fig:SFR}).} 
 \label{fig:SSFRcorrs}
\end{figure}

%[discuss specific SFRs and time-scales, and plot Fig.~\ref{fig:SSFRcorrs}.]
We end by discussing the specific SFR, defined as ${\rm sSFR}\equiv{\rm SFR}/M_\ast$. 
The sSFR is a useful physical quantity, because by normalizing by the stellar mass, relations with star formation activity can appear clearer.  
In Figure~\ref{fig:SSFRcorrs}, we show the distribution of sSFR versus dust temperature and the dust/stellar luminosity ratio for pixels in the LMC and SMC, analogous to Figure~\ref{fig:SFRcorrs} (and to {Fig.~7} in S11, for whole galaxies). We see that the LMC has a wider range of sSFRs than the SMC, and that the weak correlation between SFR and $T_\mathrm{dust}$ disappears when accounting for $M_\ast$. 
However, the correlation between sSFR and $L_\mathrm{dust}/L_\ast$ is strong for both galaxies, consistent with S11: regions with larger dust/stellar ratios tend to have higher sSFRs. 
In addition, the sSFR defines a characteristic time-scale for star formation: $\tau_{\rm SFR}^{-1}=\dot{M}_\ast/M_\ast$ (Brinchmann et al.\ 2004).  This implies that dusty star-forming regions, such as 30 Dor and the SMC wing (see Fig.~\ref{fig:LdustLstar}), have short star formation time-scales, on the order of a Gyr.
% comparable to the gas consumption time-scale, ${\rm SFE}_t^{-1}$, above.

\section{Summary and Discussion}\label{sec:discussion}

We now summarize our main conclusions: 

\begin{itemize}
 \item Utilizing the entire dust SED, including \textit{Spitzer} (SAGE) and \textit{Herschel} (HERITAGE) observations, we have spatially resolved dust luminosities throughout the LMC and SMC.  
 %We do not detect a substantial cold dust component in either galaxy.  
 Some regions with bright $L_\mathrm{dust}$ coincide with known star-forming regions.
 \item Dust temperature is more strongly correlated with $L_\mathrm{dust}/M_\mathrm{dust}$ than $L_\mathrm{dust}$ alone.  In addition, we detect a significant 70~$\mu$m excess (with respect to modified-blackbody fits at longer wavelengths) in both galaxies, especially the SMC, indicating the presence of stochastically heated dust and/or that the dust is heated to a range of temperatures because of a range of starlight intensities. 
 \item Dust/stellar luminosity and mass ratios exhibit substantial variation throughout both the LMC and SMC.  Bright star-forming regions (with concomitant dust emission) have bright $L_\mathrm{dust}/L_\ast$, while regions with many stars, especially the LMC bar, have low dust/stellar ratios.
 \item We probe the spatial distribution of star formation in the LMC and SMC, using $H\alpha$ emission, and find that regions with high SFR correlate with those with warm dust temperature and bright $L_\mathrm{dust}/L_\ast$. 
 \item We compare `global' properties of the LMC and SMC to other galaxies (e.g., Skibba et al.\ 2011).  The dust/stellar mass ratios, %($6.2\times10^{-4}$ and $3.0\times10^{-4}$ for the LMC and SMC) 
especially of the SMC, are lower than that of similar 
metal-poor dwarf and late-type galaxies.  The LMC approximately lies on the stellar mass-SFR relation of disk galaxies, but it has a particularly large star formation efficiency (quantified by $\mathrm{SFR}/M_\mathrm{gas}$ or $M_\ast/(M_\ast+M_\mathrm{gas})$) compared to other galaxies.
\end{itemize}

%%[discuss \textbf{implications of our results for galaxy evolution}.  
%%metallicity and environmental dependences? SMC/LMC tidal interaction? connection between dust production/destruction and SF history?] %mention the different SFE's too?
%As discussed in Section~\ref{sec:diffs}, many of the dust and stellar 
%properties of the Magellanic Clouds are consistent with what one might expect 
%based on the galaxies' metallicities, stellar masses, and morphologies. 
%Nonetheless, the LMC's high star formation efficiency and the SMC's 
%relatively high SFR density may be somehow due to their unique formation 
%histories, including tidal stripping and triggered star formation.  

The LMC and SMC appear to be a pair of interacting and infalling satellite 
galaxies (perhaps on their first passage through the MW halo), which is a 
relatively rare phenomenon (Liu et al.\ 2011; Besla et al.\ 2012). 
It is an interesting question to ask, to what extent these environmental 
processes (interaction and infall) are connected with the galaxies' evolution, 
especially their dust production/destruction and star formation histories.
It is well known that tidal encounters or `harassment' between close pairs of 
galaxies can modify their morphologies and induce bursts of star formation 
(Moore et al.\ 1996; Barton et al.\ 2000), 
but the LMC and SMC allow a close examination of the effects of these processes, 
including the indirect effects on dust heating and emission (both in 
star-forming regions and the diffuse ISM).
For example, the LMC has essentially become a one-armed spiral galaxy, in which the 
arm has ongoing star formation and substantial dust emission; its off-center bar has 
limited star formation with a relatively warm dust temperature.
% Jacco: could comment on fact that LMC's SFR map doesn't have spiral structure (like the stellar mass sort of does).

%[also discuss how these results can be used to \textbf{constrain models, simulations, and SED templates}. 
%how are (global or local) galaxy properties related to dust \& stellar SEDs?]
%maybe also mention high-z studies?
Finally, our results for the global, and especially resolved, galaxy 
properties of the Magellanic Clouds can potentially be used to constrain 
models, simulations, and SED templates. 
Semi-analytic models of galaxy formation make assumptions about the energy 
balance and spatial distribution of stars and dust (e.g., Lacey et al.\ 2008; 
Somerville et al.\ 2012), while our results constrain the strength and 
scatter in the correlations between dust luminosity (or $L_\mathrm{dust}/L_\ast$) 
and dust temperature and SFR (Sections~\ref{sec:dusttostellar} and \ref{sec:sfr}). 
% maybe mention SED assumptions in gal form models: e.g., Lacey et al. 2008; Somerville et al. 2012
%
%[comment on emission from diffuse `cirrus' component vs stellar `birth clouds', and possible constraints on RT models.] %(Jonsson et al.\ 2010; Narayanan et al.\ 2010)
Furthermore, radiative transfer models have already been compared to the SEDs of 
nearby galaxies (e.g., Misselt et al.\ 2001; Jonsson et al.\ 2010; Silva 
et al.\ 2011) and could benefit from the additional constraints on dust 
properties provided by \textit{Herschel} (see Section~\ref{sec:results}), 
such as the distributions of dust luminosity, temperature, and mass. 

\subsection{Differences between the Magellanic Clouds}\label{sec:diffs}

%\textbf{move to discussion?}
%Chad: How do the clouds differ, and to what can the differences be attributed (e.g., mass, composition)?

Though the Magellanic Clouds are often studied together, they have many differences. 
The SMC is more dwarf-like than the LMC, which is classified as a barred one-armed spiral galaxy (de Vaucouleurs \& Freeman 1972), %de Vaucouleurs G., Freeman K. C., 1972, Vistas in Astronomy, 14, 163     ...and by RC3 (de Vaucouleurs et al.\ 1991)
and has a much lower metallicity than the LMC (Russel \& Dopita 1992).  
%
% mention other differences not studied here, such as gas content and specific stellar populations...
% e.g., SMC (and tail) has larger gas/dust than LMC, and both are larger than MW's.
The MCs also have different gas/dust ratios (Gordon et al.\ 2003; Leroy et al.\ 2011), 
submillimeter excesses (Bot et al.\ 2010; Israel et al.\ 2010), 70~$\mu$m excesses (Appendix~\ref{app:70umexcess}), 
dust production from evolved stars (Srinivasan et al.\ 2009; Boyer et al.\ 2012; Riebel et al.\ 2012), 
%Riebel D., Srinivasan S., Sargent B., Meixner M., 2012, ApJ, 753, 71
star formation histories (Harris \& Zaritsky 2004, 2009), and 
proper motions (Kallivayalil et al.\ 2006). %and Besla et al. 2007

In Section~\ref{sec:dusttostellar}, we found that the SMC has a 
%higher 70$\mu$m excess and a 
slightly lower dust/stellar luminosity and mass ratio %(but similar dust/stellar mass) 
% also SFR (or SFR/$M_\ast$) difference?
than the LMC, on average.  
%(They have similar dust mass and luminosity distributions, however, the LMC's total dust and stellar mass is larger.)
%
The SMC's stellar and dust masses are five and ten times lower than the LMC's, respectively. 
The SMC also has a much lower SFR and a lower star formation efficiency (SFE) than the LMC. 

%\textbf{[this needs further discussion.]} 
Many of these different properties of the galaxies could be simply explained by the fact that 
the SMC is considerably less massive and more metal-poor than the LMC. 
For example, less massive galaxies are expected to have lower SFRs than more massive ones, and metal-poor galaxies are expected to have higher gas/dust mass ratios.
Nonetheless, the SMC's higher SFR density (Fig.~\ref{fig:SFRcorrs}) and the LMC's relatively high SFE, 
%possibly due to tidally triggered star formation, 
may be the result of these galaxies' unique formation histories, including tidal stripping and triggered star formation. 
%As discussed in Section~\ref{sec:diffs}, many of the dust and stellar 
%properties of the Magellanic Clouds are consistent with what one might expect 
%based on the galaxies' metallicities, stellar masses, and morphologies. 
%Nonetheless, the LMC's high star formation efficiency and the SMC's 
%relatively high SFR density may be somehow due to their unique formation 
%histories, including tidal stripping and triggered star formation.  

\subsection{Selected Regions in the MCs}\label{sec:regions}%{Some `Unusual' Regions}

% label some regions in Fig~\ref{fig:MLdust}.  LMC: bar, 30 Dor, N11.  SMC: "tail".
% Chad, on labeling regions with DS9: You should be able to do it with a region file, since there's a text mode.  There aren't a ton of fonts available, but it should be possible to make something presentable.
We now discuss a few selected regions: the bar and two well-known star-forming regions, 30~Dor and N11, in the LMC, and the SMC wing. %(and bridge?)
(These regions are labeled in Fig.~\ref{fig:MLdust}.) 
% I already have the right selection (ra,dec) for N11 somewhere.  see Tfit on marvin & einstein.

The LMC's bar has a large stellar luminosity and mass, indicated by bright near-IR luminosities. 
%by its low dust/stellar ratios (Figs.~\ref{fig:LdustLstar} and \ref{fig:MdustMstar}).  
It has a large fraction of old stars, with ages $>10~\mathrm{Gyr}$ (Harris \& Zaritsky 2009). 
In addition, its dust mass/luminosity ratio is relatively low (Fig.~\ref{fig:MLdust}) 
and dust temperature is relatively high; considering its limited recent star formation, 
this suggests that the dust is heated by the ISRF. %interstellar radiation field. 
% This is reminiscent of the bulge components in other galaxies (Zaritsky 2004; Skibba et al. 2011), but there is evidence against the `bar' being a half-obscured bulge (H\&Z 2009). 
Interestingly, the LMC's photometric center, which is centered on the bar, is offset by $\sim1~\mathrm{kpc}$ from the stellar and H\,{\sc i} kinematic center (Cole et al.\ 2005). %, and the bar's dust emission appears offset as well (Figs.~\ref{fig:Ldust} and \ref{fig:LIRratio}).  
Unlike some barred spiral galaxies (see the evolving bar fraction in Sheth et al.\ 2008), the LMC's bar has been an integral part of the galaxy for most of its history (Harris \& Zaritsky 2009), and this asymmetry in the bar appears to have been the result of a recent collision with the SMC (Bekki \& Chiba 2007).

There are large H\,{\sc ii} region complexes near the LMC's bar, such as 30~Dor and N44 %also N79/N83
as well as others further away, such as N11 in the northwest. %(Indebetouw et al.\ 2009; Carlson et al.\ 2012). 
These regions have substantial stellar emission as well as dust emission, though this is not 
accompanied by a particularly large dust mass (i.e., they have low $M_\mathrm{dust}/L_\mathrm{dust}$; see Fig.~\ref{fig:MLdust}). 
This suggests that these regions have a relative underabundance of dust grains, 
possibly due to dust destruction by stellar winds or %supernova
shocks (Jones et al.\ 1994; Paradis et al.\ 2009); however, these star-forming regions, especially 30~Dor, have enhanced dust/gas ratios (Paradis et al.\ 2011), indicating that most of their gas has been consumed.
% need more discussion: see Frank's comments.

As discussed above, the SMC wing and bridge appear to be tidally stripped and their star formation 
tidally triggered (Harris 2007; Gordon et al.\ 2009, 2011). 
In addition, the wing has a high gas/dust mass ratio, possibly due to dust destruction 
by a harder radiation field and shocks during the tidal interaction (Gordon et al.\ 2009). 
We find that the wing also has high dust/stellar ratios (Figs.~\ref{fig:LdustLstar} and \ref{fig:MdustMstar}), 
but these are likely due to star formation, not to a significant dust abundance, 
making the wing somewhat similar to H\,{\sc ii} regions within the MCs.

%\vspace{0.2cm}
\section*{Acknowledgments}

% Acknowledge Gonzalo or include him as co-author.
% thank people for discussions, like Dennis Zaritsky. maybe talk to Kelly too.
RAS acknowledges financial support from the NASA \textit{Herschel} Science Center, JPL contract \#1350371.  
We also acknowledge support from the NASA \textit{Herschel} Science Center, JPL contracts \#1381522 and \#1381650. 
AL is supported in part by NSF AST-1109039. 
MR acknowledges support from CONICYT through FONDECT grant No.\ 1080335. 
%We thank Gonzalo Aniano for kindly providing his convolution kernels (Aniano et al.\ 2011). 
We thank the anonymous referee for insightful comments that helped to improve the quality of the paper. %comments and recommendations?  improved presentation of the results? 
We thank the contributions and support from the European Space Agency (ESA), the PACS 
and SPIRE teams, the \textit{Herschel} Science Center, and the NASA \textit{Herschel} 
Science Center (esp.\ A.\ Barbar and K.\ Xu) and the PACS and SPIRE instrument control 
centers, without which none of this work would be possible.

%We thank XXX for valuable discussions about our results.
% Maybe show the completed draft to a few others, like Joannah, Kelly, and Dennis Z.  
%other potential people with interesting advice: Sharon M. and Stefano, maybe Lars Mattsson.

%HERITAGE acknowledgment. 
%Herschel is an ESA space observatory with science instruments provided by European-led Principal Investigator consortia and with important participation from NASA. SPIRE has been developed by a consortium of institutes led by Cardiff University (UK) and including Univ. Lethbridge (Canada); NAOC (China); CEA, LAM (France); IFSI, Univ. Padua (Italy); IAC (Spain); Stockholm Observatory (Sweden); Imperial College London, RAL, UCL-MSSL, UKATC, Univ. Sussex (UK); and Caltech, JPL, NHSC, Univ. Colorado (USA). This development has been supported by national funding agencies: CSA (Canada); NAOC (China); CEA, CNES, CNRS (France); ASI (Italy); MCINN (Spain); SNSB (Sweden); STFC (UK); and NASA (USA).

%%%%% means that the reference was added in the new draft.  maybe all these refs aren't necessary.

%\appendix

\begin{appendix}

\section{Total Infrared Luminosity}\label{app:LTIR}

%We begin by analyzing the total infrared (TIR) luminosity of the Magellanic Clouds. 
%The TIR luminosity is a useful quantity because it can be 
%directly inferred from the IR fluxes, and because it can be used as a proxy 
%for the obscured star formation as well as the temperature of dust grains 
%(e.g., Dale \& Helou 2002, Draine \& Li 2007).  
We follow Draine \& Li (2007), and use a common calibration of the total IR luminosity of dust, 
% i.e., out to $1000\mu$m wavelengths
inferred from 8, 24, 70, and 160~$\mu$m photometry: 
\begin{equation}
  L_\mathrm{TIR}\,=\,0.95\langle \nu L_\nu\rangle_{8.0} + 1.15\langle\nu L_\nu\rangle_{24} + \langle\nu L_\nu\rangle_{70} + \langle\nu L_\nu\rangle_{160}
 \label{LTIR}
\end{equation}
where $\langle\nu L_\nu\rangle\equiv\langle L_\nu\rangle c/\lambda$. 
We have also tested the Dale \& Helou (2002) formula, which uses only the 
MIPS bands, and have obtained very similar results. 
Note that with this definition, $L_\mathrm{TIR}$ is inferred from wavelengths of $\lambda\leq160~\mu$m, in contrast with $L_\mathrm{dust}$ (Eqn.~\ref{eq:fdust}), which is summed over the entire dust SED.

\begin{figure*}
 \figurenum{A1}
 %\centerline{\includegraphics[width=0.497\hsize]{LMC_LdustLTIR_new_Karlnanbg.png}} %{LMC_LTIRLdust_Karlnanbg.png} %{LMC_LTIRLdust_ratio_new.png} 
 \includegraphics[width=0.497\hsize]{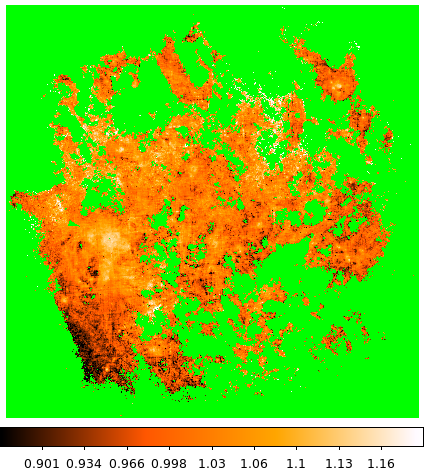}%{LMC_LdustLTIR_new_Karlnanbg.png} 
 \includegraphics[width=0.497\hsize]{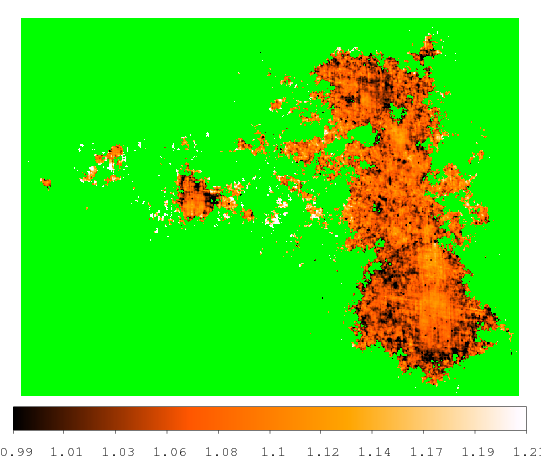}%{SMC_LdustLTIR_new_Karlnanbg.png} 
 \caption{LMC (left) and SMC (right) (left) and SMC (right) (left) and SMC (right) (left) and SMC (right) (left) and SMC (right) (left) and SMC (right) (left) and SMC (right) (left) and SMC (right) $L_\mathrm{dust}/L_\mathrm{TIR}$ ratio.  $L_\mathrm{dust}$ includes the entire dust SED, while $L_\mathrm{TIR}$ infers it from shorter wavelengths ($\lambda\leq160~\mu$m).}
 \label{fig:LIRratio}
\end{figure*}

We show the ratio of the dust luminosity $L_\mathrm{dust}$, which includes longer wavelength data, to this $L_\mathrm{TIR}$ in Figure~\ref{fig:LIRratio}. 
%
%their ratio is interesting too, since LTIR is inferred from wavelengths of $\lambda\leq160\mu$m.
It is interesting that the $L_\mathrm{TIR}$ and $L_\mathrm{dust}$ maps are similar, but not identical.  Their ratio is not uniformly unity; for example, it varies in the SE region of the LMC, and is relatively large in the bright star-forming region 30~Dor. 
This is consistent with ongoing work (Galametz et al., in prep.), showing some differences between IR luminosities inferred from \textit{Spitzer} bands ($\lambda\leq160~\mu$m) and \textit{Herschel} bands in regions with very high or low surface brightnesses, though in general the submm contribution is relatively small in low-metallicity galaxies. 
%For the SMC, $L_\mathrm{dust}/L_\mathrm{TIR}$ is more uniform, but the ratio has typical values of  $0.71\pm0.05$ for both MCs.
The $L_\mathrm{dust}/L_\mathrm{TIR}$ ratio of the pixels in the maps has a mean of 
$1.02\pm0.0001$ (with standard deviation $\sigma=0.06$) in the LMC and $1.07\pm0.0002$ (with $\sigma=0.05$) in the SMC. 
%, which is statistically consistent with unity. 
%Considering that $L_\mathrm{dust}$ includes SPIRE photometry at far-IR and submm wavelengths--fluxes inferred from shorter wavelengths by the $L_\mathrm{TIR}$ calibration---these ratios suggest that there may be a significant cold dust component emitting at submm wavelengths. 
%Such emission may be related to the `submillimeter excess' that has been observed among nearby dwarf galaxies (Bot et al.\ 2010; Galametz et al.\ 2009, 2011; Gordon et al.\ 2010; Galliano et al.\ 2011), but not among typical nearby late-type galaxies (Dale et al.\ 2012). 
%Nonetheless, it is simply possible that, by utilizing the entire dust SED, $L_\mathrm{dust}$ more accurately accounts for the spatial variations of dust heating, which is explored further in Section~\ref{sec:Tdust}. 
The $L_\mathrm{TIR}$ luminosity inferred from shorter wavelengths is approximately accurate, 
but estimates based on these wavelengths alone could be slightly biased by different grain size distributions or emissivities. %or even PAH abundances. 
%Frederic: ...you would encounter in the LMC SEDs hotter, for instance, or different size distributions that would bias the 24 μm coefficient, or different PAH fractions that would bias the 8 μm coefficient, or different grain emissivities biasing the 160 μm coefficient.
%
%indicating that these galaxies do not harbor a substantial cold dust component (which would result in $L_\mathrm{dust}\gg L_\mathrm{TIR}$).  Nonetheless, 
By utilizing the entire dust SED, $L_\mathrm{dust}$ more accurately accounts for the spatial variations of emission from small and large dust grains. 
%of dust heating, which is explored further in Section~\ref{sec:Tdust}.

\section{70 micron Excess}\label{app:70umexcess}

\begin{figure*}
 \figurenum{B1}
 %plot on ds9 using "histogram" rather than "linear"?
 \includegraphics[width=0.497\hsize]{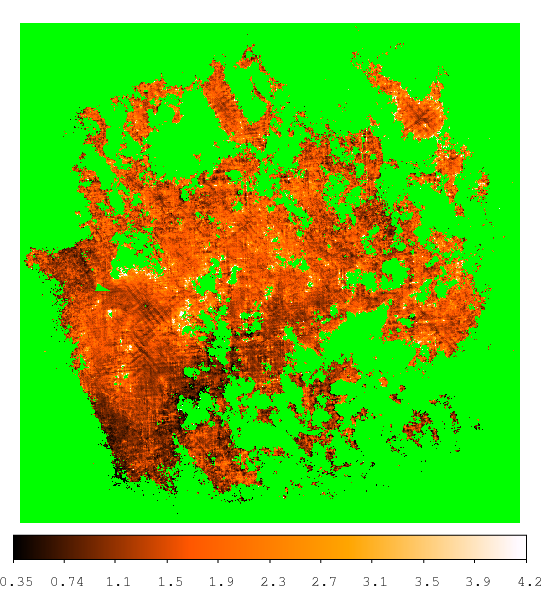}%{LMC_mips70predicted70um_Karlnanbg.png} 
 \includegraphics[width=0.497\hsize]{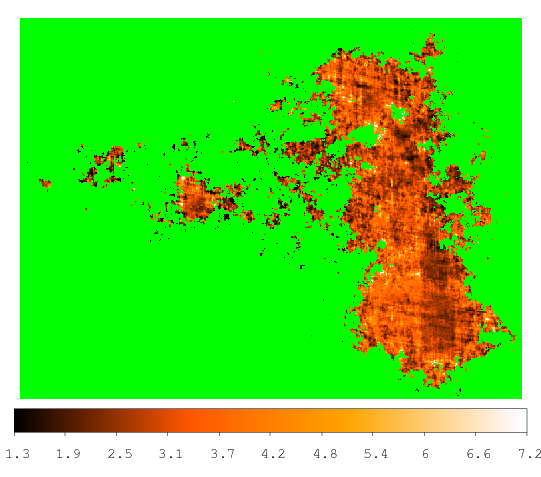}%{SMC_mips70predicted70um_Karlnanbg.png} 
 \caption{LMC (left) and SMC (right) 70~$\mu$m excess, $f_{70{\rm, MIPS}}/f_{70{\rm, modBB}}$, computed as the ratio between measured MIPS70~$\mu$m flux and 70~$\mu$m flux predicted from modified blackbody model (fitted to 100-500~$\mu$m fluxes). 
  }
 \label{fig:70umexcess}
\end{figure*}

Using the single-temperature modified blackbody fits to the 100-500~$\mu$m dust SEDs, 
we can compare the 70~$\mu$m flux predicted from this to the observed MIPS 70~$\mu$m flux. 
This ratio is an indicator of the spatial distribution of stochastically heated small dust grains 
(Draine \& Li 2001; Galliano et al.\ 2003), and/or the fact that the dust grains of different sizes are heated to a range of temperatures (rather than a single $T_{\rm dust}$) by a range of starlight intensities. 
We show maps of it for the LMC and SMC in Figure~\ref{fig:70umexcess}. 
A ratio (observed/predicted) of greater than unity is evidence for a 70~$\mu$m `excess' (Bot et al.\ 2004; Bernard et al.\ 2008).  
% from bkgdnoise.pro: LMC has 1.402\pm0.578; SMC has 2.814\pm1.006
We find that most of both MCs have a significant excess, with the LMC and SMC having 
mean excesses of approximately 1.4 (rms 0.6) and 2.8 (rms 1.0), respectively. 
%\textbf{but why doesn't Fig.~\ref{fig:70umexcess}a look like Jean-Philippe's Fig.~6?} 
The result in the left panel of Figure~\ref{fig:70umexcess} is not in disagreement with Bernard et al.\ (2008), 
who employed a different definition of the 70~$\mu$m excess, with respect to a model that includes very small grain emission, and found a strong excess in the LMC in regions to the east and south of 30~Doradus. 
%in some disagreement with Bernard et al., who found a strong excess in the LMC only in the regions to the East and South of 30~Doradus; however, their analysis was limited to MIPS and IRAS far-IR fluxes, while ours benefits from PACS and SPIRE, probing longer wavelengths.

% Jacco suggested commenting on the fact that in some regions of the LMC, the 70um flux is *under*estimated by the modBB.  How does that occur?

Since we observe $f_{70{\rm, MIPS}}/f_{70{\rm, modBB}}$ ratios with a significant excess of unity, this implies that a significant fraction of small dust grains are stochastically heated by photon absorption.  The larger ratios within the SMC is expected because the SMC has a steeper UV extinction law, which likely means that the grains in the SMC are generally smaller than those in the LMC. 
%Jacco: warmer dust in compact star-forming dust clouds in the SMC compared to those in the LMC, based on MIPS SED observations (van Loon et al.\ 2010).
\end{appendix}

%\label{lastpage}

\end{document}